\documentclass[preprint,showpacs]{revtex4}
\usepackage[dvips]{graphicx}
\usepackage{epsfig}
\usepackage{psfrag}
\usepackage{amsmath}
\usepackage{amsfonts}
\usepackage{amssymb}
\usepackage{bm}
\begin{document}
\title{Low temperature breakdown of coherent tunneling in amorphous solids
induced by the nuclear quadrupole interaction}
\author{A. L. Burin$^{\dag\ast\spadesuit}$, I. Ya. Polishchuk$^{\ddagger\spadesuit}$,
P. Fulde$^{\spadesuit}$, Y. Sereda$^{\dag}$}
\affiliation{$^{\dag}$Department of Chemistry, Tulane University, New Orleans, LA 70118}
\affiliation{$^{\ast}$William P. Fine Institute of Theoretical Physics, School of Physics
and Astronomy, }
\affiliation{University of Minnesota, Minneapolis, MN D-01187}
\affiliation{$^{\ddag}$RRC Kurchatov Institute, Kurchatov Sq. 1, 123182 Moscow, Russia}
\affiliation{$^{\spadesuit}$Max-Planck-Institut f\"ur Physik Komplexer Systeme, D-01187
Dresden, Germany}
\date{\today}

\begin{abstract}
We consider the effect of the internal nuclear quadrupole interaction on 
quantum tunneling in complex multi-atomic two-level systems. Two distinct
regimes of strong and weak interactions are found. The regimes depend on
the relationship between a characteristic energy of the nuclear quadrupole interaction
$\lambda_{\ast}$ and a bare tunneling coupling strength $\Delta_{0}$. When
$\Delta_{0}>\lambda_{\ast}$, the internal interaction is negligible and 
tunneling remains coherent determined by $\Delta_{0}$. When
$\Delta_{0}<\lambda_{\ast}$, coherent tunneling breaks down and an
effective tunneling amplitude decreases by an exponentially small overlap
factor $\eta^{\ast}\ll1$ between internal ground states of left and
right wells of a tunneling system. This affects thermal and kinetic properties of tunneling systems 
at low temperatures $T<\lambda_{*}$.
The theory is applied for interpreting the anomalous behavior of the resonant
dielectric susceptibility in amorphous solids at low temperatures $T\leq 5$mK where the
nuclear quadrupole interaction breaks down coherent tunneling. We suggest the experiments with external
magnetic fields to test our predictions and to clarify the internal
structure of tunneling systems in amorphous solids.
\end{abstract}

\maketitle

\section{Introduction}

\label{intr}

A transition between two energy minima separated by a potential barrier $U$
occurs in different ways at high and low temperatures. At high temperatures
the motion between the energy minima is classical, i.e., thermally activated,
suggesting that the particle acquires the energy $U$ from the environment to
overcome the barrier. This results in an exponential Arrhenius factor
$\exp(-U/T)$ for the transition rate. The classical transition rate decreases
rapidly with decreasing the temperature. At low temperature the
above-described transport crosses over into the weakly temperature dependent
quantum tunneling. A tunneling transition rate is essentially governed by the
exponentially small and temperature--independent factor $\exp(-2S(U)/\hbar)$,
where $S(U)$ describes the classical action for the motion through the
inverted potential barrier. The strong sensitivity of the tunneling exponent
against particle mass makes quantum tunneling more favorable for light
particles as electrons, while the motion of heavy nuclei is classical down to
temperatures of order of Kelvin. However, for $T\leq 1$K quantum tunneling
displays itself in the thermodynamic and kinetic properties of atomic systems.
One impressive example is amorphous solids, in which the low temperature
properties are governed by the two-level systems (TLS's). These TLS's are made
of atoms or groups of atoms, experiencing a tunneling motion between pairs
of energy minima separated by potential barriers \cite{ahvp}. They contribute
to the universal thermodynamic and kinetic properties in all known glasses as
well as some other disordered materials for $T\leq1K$
(see Refs.  \cite{Zeller,Hunklinger,Phillips_review}).

A TLS is described by the standard pseudospin 1/2 Hamiltonian 
\begin{equation}
\widehat{h}_{TLS} = -\Delta_{0} \cdot s^{x} - \Delta\cdot s^{z}.
\label{eq:hamtls}
\end{equation}
Here $\Delta_{0}$ is a tunneling amplitude coupling two energy minima and
$\Delta$ is a level asymmetry. The quantity $\Delta_{0}$ is exponentially
sensitive to external and internal parameters of the system. This results in a
logarithmically uniform distribution of tunneling amplitudes for various TLS's
\begin{equation}
P(\Delta, \Delta_{0}) = \frac{P}{\Delta_{0}}, ~~P = const. \label{Eq:TLSmain}
\end{equation}
This distribution of TLS's leads to the universal temperature and time dependencies for 
various physical characteristics of amorphous solids, thus making
"glassy\textquotedblright\ behavior very easily recognizable. For instance,
this includes the logarithmic relaxation in time of the specific heat
\cite{ahvp,Zeller,Hunklinger,Phillips_review} and the non-equilibrium dielectric
susceptibility \cite{DDO1}. In
addition, dielectric and acoustic (a sound velocity) susceptibilities in glasses
show a logarithmic temperature dependence also associated with the
distribution Eq. (\ref{Eq:TLSmain}). Note that the exponential sensitivity of
a tunneling strength $\Delta_{0}$ to environmental interactions also
broadens noticeably the distribution of tunneling amplitudes and relaxation
rates in other systems, e.g., tunneling of a large electronic spin in magnetic
molecule Mn$_{12}$Ac. The latter system shows a broad spectrum of relaxation times
due to interaction of electronic spins with nuclear spins \cite{Barbara,Stamp}. A broad spectrum of relaxation times also has been reported in new
disordered magnetic alloys \cite{Rosenbaum}.

The nature of the tunneling systems in amorphous solids remains unclear in
spite of the theoretical efforts attempting various models
\cite{Klinger,Parshin,Leggett,Burin_Kagan,b6}. The main problem in the theory
is a lack of experiments which are capable of proving the advantage of a
specific model against the original phenomenological model \cite{ahvp} based
on Eq. (\ref{Eq:TLSmain}). The phenomenological
approach Eq. (\ref{Eq:TLSmain}) can be used to explain a variety of experimental
data, while the interaction between TLS can successfully be treated as a weak
correction \cite{BurinJLTP,b6}.
Even the internal TLS structure is unclear yet. For instance,
nobody knows how many atoms do participate in a single tunneling event. We
hope that understanding the nuclear quadrupole interaction effects in glasses, which are 
sensitive to the internal
TLS structure, will help to resolve this question. 

Recent experimental investigations of amorphous solids at very low
temperatures have revealed a number of qualitative deviations from the
predictions of the standard tunneling model Eq. (\ref{Eq:TLSmain}). In
particular, it is demonstrated in several works \cite{SEH,DDO2,b6} that, for
$T\leq 5$mK, the expected logarithmic temperature dependence of the dielectric
constant breaks down and the dielectric constant becomes approximately
temperature-independent. This result conflicts with the logarithmically
uniform distribution Eq. (\ref{Eq:TLSmain}) of TLS's over their tunneling
amplitudes. To resolve the problem, one can assume that the distribution of
TLS's has a low-energy cutoff at $\Delta_{0,min}\simeq5$mK. This assumption,
however, contradicts the observation of very long relaxation times in all
glasses. These times (a week or longer) require much smaller tunneling
amplitudes \cite{DDO1} than $5$mK (remember that the TLS relaxation time is
inversely proportional to its squared tunneling amplitude).

We suggest the explanation of this controversial fact by using the recent model
of W\"{u}rger, Fleischmann, and Enss \cite{WFE} who proposed that the nuclear
quadrupole interaction affects the properties of TLS's at low temperatures by the 
mismatch of the nuclear quadrupole states in different potential wells (see
Fig. \ref{fig1}). This is very similar to the electronic spin tunneling
suppression by the nuclear spin interaction in magnetic molecules \cite{Stamp}. The
significance of the nuclear quadrupole interaction has recently been proven
experimentally in glycerol glass \cite{D1}. This interaction helps to
understand the anomalous magnetic field dependence of
dielectric properties in entirely non-magnetic dielectric glasses
\cite{EnssReview,Ludwig,SEH}.

We show that an effective tunneling amplitude of TLS having an energy less than  
its nuclear quadrupole interaction  is remarkably reduced due to the 
mismatch of TLS nuclear quadrupole states in its two energy minima, similarly to the polaron effect. Consequently 
the spectrum of TLS tunneling amplitudes $\Delta_{0}$ possesses a pseudogap below the nuclear quadrupole 
interaction energy $\lambda_{*}$. Dielectric and acoustic susceptibilities of glasses 
become temperature indepedent for temperatures belonging 
to this pseudogap because there is no TLS with $\Delta_{0}\sim T$ to contribute. 
Thus our model explains the experimental observations. 
In addition, we predict that the application of a strong external magnetic 
field will reduce the mismatch of different nuclear quadrupole states and thus 
restore the logarithmic temperature dependence of TLS susceptibilities. 
According to our theory dielectric glasses having no nuclear quadrupole interaction should obey the 
predictions of the tunneling model, which mostly agrees with the experiment.

Since a low temperature dielectric constant and a speed of sound in
glasses have a similar physical nature they should have
a similar behavior at low temperatures. Therefore the saturation in the
logarithmic temperature dependence of a sound velocity should be seen at
$T<10$mK in materials possessing non-vanishing nuclear quadrupole moment. The
situation with sound velocity measurements is, however, more
complicated because it is more difficult to perform the low-temperature
measurements for a sound velocity then for a dielectric constant. There
exists a few measurements that we discuss together with the measurements of
the dielectric constant. The present theory is developed for the dielectric
constant, although our conclusions can be extended to the velocity of sound
without major changes.

The paper is organized as follows. In Section \ref{model_and_properties} the
model of two-level systems affected by the nuclear quadrupole interaction is
introduced. Then, we discuss qualitatively 
the renormalization of tunneling amplitude by the nuclear quadrupole interaction 
depending on the
relation between tinneling splitting $\Delta_{0}$ and a quadrupole
interaction $\lambda_{*}$ . In Section \ref{eps} the expression for the
resonant dielectric susceptibility is obtained and the influence 
on the tunneling amplitude renormalization on the resonant part of the TLS
dielectric constant is described in the qualitative level. In Section \ref{pert_theory} we use a
perturbation theory to characterize quantitatively the temperature dependence of the
dielectric constant in the presence of nuclear quadrupole interactions in the high and 
low temperature limits. In
Section \ref{eq:oscill} we consider a solvable model of
TLS's coupled to harmonic oscillators, conveniently replacing nuclear spins.
The solution of this problem permits us to simulate the influence of the
nuclear quadrupole interaction on a TLS dielectric constant at all temperatures of interest. 
 In Section \ref{sect:h} the effect of the external
magnetic field on the dielectric constant is considered within the simplified
model of Sect. \ref{eq:oscill}. In Section \ref{exp_params} the parameters of our model are compared with
the experimental data. In final Section \ref{sect:disc} the conclusions are formulated and the
suggestions for an experimental verification of our theory are made. 
The short version of the manuscript appears 
in the Physical Review Letters \cite{prl2005}.

\bigskip

\section{Model}

\label{model_and_properties}

\subsection{Nuclear Quadrupole Interaction}

\label{H}

How does the nuclear quadrupole interaction affect tunneling? Consider a
tunneling system formed by $n$ atoms all possessing a nuclear spin $I\geq1$ and
consequently a nuclear electrical quadrupole moment.The total tunneling
Hamiltonian $\widehat{h}$ can be described by the standard TLS pseudospin
Hamiltonian $\widehat{h}_{TLS}$ (\ref{eq:hamtls}) and the quadrupole interactions
$\widehat{H}_{r}$ in the right well ($s^{z}=1/2$) and $\widehat{H}_{l}$ in the
left well ($s^{z}=-1/2$) as follows
\begin{equation}
\widehat{h}=\widehat{h}_{TLS}+\frac{\widehat{H}_{r}+\widehat{H}_{l}}
{2}+(\widehat{H}_{r}-\widehat{H}_{l})s^{z}. \label{eq:tls_Ham}
\end{equation}
The local nuclear quadrupole Hamiltonians $\widehat{H}_{r,l}$ can be expressed
as a sum of interactions of all $n$ nuclear spins $\mathbf{{\widehat
{I}_{i}}}$ over all $n$
atoms that simultaneously participate in its tunneling motion 
with the local electric field gradient tensors $\partial
F_{a}^{(l,r)}/\partial x_{b}$ different in general for the right and left wells
\begin{align}
\widehat{H}_{r,l}  &  =\sum_{i=1}^{n}\widehat{h}_{i}^{(r,l)},\nonumber\\
\widehat{h}_{i}^{(r,l)}  &  =-\frac{Q}{2}\sum_{a,b=x,y,z}\left(  \widehat
{I}_{i}^{a}\widehat{I}_{i}^{b}+\widehat{I}_{i}^{b}\widehat{I}_{i}^{a}
-2\delta_{ab}\frac{I(I+1)}{3}\right)  \frac{\partial F_{a}^{(l,r)}}{\partial
x_{b}}. \label{eq:Quadr_int}
\end{align}
Here $\widehat{I}_{i}^{a}$ is a nuclear spin projection onto the $a-$axis and
$Q$ is the electrical quadrupole moment. 

\begin{figure}
[ptbh]
\begin{center}
\includegraphics[
height=1.3759in,
width=2.3609in
]
{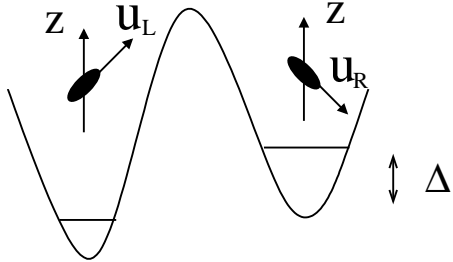}
\caption{The two level system having different nuclear
quadrupole quantization axes $u_{L}$ and $u_{R}$ defined by the local electric
field gradient in the left and right wells. }%
\label{fig1}
\end{center}
\end{figure}

In what follows, we consider a simplified model for the nuclear quadrupole
interaction (\ref{eq:Quadr_int}) possessing the axial symmetry(see Fig.
\ref{fig1})
\begin{equation}
\widehat{H}_{l,r}=b\left(  \left(  I_{l,r}^{~u_{l,r}}\right)  ^{2}
-\frac{I(I+1)}{3}\right). \label{eq:ax_sym_mod}
\end{equation}
Here 
\begin{equation}
b\approx Q\mid\partial F/\partial x\mid, \label{eq:Quadr_int1}
\end{equation}
and $u_{L}$ and $u_{R}$ define the directions of the electric field gradient
in the left and right wells, respectively.

There exists also the magnetic-dipolar interaction of nuclear spins. It is
usually by several orders of the magnitude smaller than the quadrupole
interaction and can be neglected.

\subsection{TLS ground state and its effective tunneling amplitude}

\label{TLS_ground_state}

It will be shown below that the resonant dielectric susceptibility of a TLS
ensemble is due to the ground state of tunneling systems affected by the
external electrical field. Only tunneling systems for which $\Delta
_{0}\sim\Delta$ can contribute to the dielectric constant. These two
parameters must exceed the thermal energy, i.e., $\Delta_{0}\geq T$ (see
\cite{Hunklinger,Phillips_review} and Sect. \ref{eps} for details). The
structure of the ground state of the Hamiltonian (\ref{eq:tls_Ham}) depends on
the relation between typical tunneling amplitude $\Delta_{0}\sim T$ and
characteristic value of the quadrupole interaction $nb$.

If $\Delta_{0}>nb,$ the nuclear quadrupole interaction can be treated as a
small perturbation and the TLS behavior obeys the standard tunneling model.

In the opposite case $\Delta_{0}<nb$, the tunneling term can be treated as a
weak perturbation. In a zero-order perturbation theory the states in the left
and right well can be considered separately. These states form a proper basis
for eigenstates of a tunneling system. In particular, the actual ground
state is a linear superposition of the ground states in the isolated left and
right wells. The energies corresponding to these ground states are
$E_{gl}=-\Delta/2+E_{g,l},$ $E_{gr}=\Delta/2+E_{g,r},$ where $E_{g,r},$
$E_{g,l}$ are the ground state energies of the quadrupole interaction
Hamiltonians $\widehat{H}_{r,l}$ (Eq.(\ref{eq:tls_Ham})) in the right or left
wells, respectively.

The effective
tunneling amplitude $\Delta_{0*}$, which couples any pair of levels in the left and right
wells of TLS's with $\Delta_{0}<nb$ decreases due to a mismatch of nuclear quadrupole ground
states in two potential wells. Obviously, when $H_{r}=H_{l},$ the nuclear
quadrupole interaction does not affect tunneling. This mismatch can be
expressed in terms of a characteristic overlap integral between the nuclear
spin ground states in the two wells. In the absence of tunneling one has
\begin{equation}
\eta_{\ast}=\eta^{n},~~\eta=\langle lg\mid rg\rangle,
\label{eq:overlpap}
\end{equation}
where the symbols $|lg>(|rg>)$ stand for the ground-state wavefunction of the
nuclear spin of a single tunneling atom in the left (right) well. This overlap
integral enters directly into the effective tunneling amplitude. In fact, at
very low temperatures and small tunneling amplitude $\Delta_{0}<nb$ one may
treat the TLS in terms of the ground states in the right and left wells. The
tunneling matrix element $\Delta_{0\ast}$ between these two states resulting
from the perturbation $-\Delta_{0}s^{x}$ is%
\begin{equation}
\Delta_{0\ast}=\eta_{\ast}\Delta_{0}. \label{eq:eff_tunnel}%
\end{equation}
Thus, the overlap integral $\eta_{\ast}$ determines the reduction of the
tunneling amplitudes due to the nuclear quadrupole interaction.

The specific value of the overlap integral depends on the absolute value of
the nuclear spin, nature of the nuclear quadrupole interaction, the number of
tunneling atoms per one TLS, and the difference in the electric field
gradients in the right and left wells. Currently, reliable information about
all these parameters is not available. In principle the value of the overlap integral $\eta$ 
fluctuates from atom to atom. This will lead to the log-normal distribution of the overall overlap integral 
$\eta_{*}$ in the large $n$ limit. As we will see below (e. g. Eq. (\ref{11})) 
the behavior of the dielectric constant is mostly sensitive to 
$\ln(\eta_{*})$, which possesses the gaussian distribution, so we can use its average value with the logarithmic accuracy of our consideration.

To estimate typical value $\eta_{\ast}$ let us turn to a simplified model
for the nuclear quadrupole interaction Eqs. (\ref{eq:Quadr_int}),
(\ref{eq:Quadr_int1}), possessing the axial symmetry. Let $\phi$ be the angle
between the directions of the electrical field gradient in the left and right wells.

First, consider the case of an integer nuclear spin. If the interaction
constant is positive, i.e., $b>0,$ the energy minimum is obtained when the
projection $I_{l,r}^{\alpha}=0$. In this case the ground state is
non-degenerate. This simplifies our consideration and makes it possible to
calculate the overlap integral of the ground state wavefunctions
depending on the mismatch angle $\phi$ (see Fig. \ref{fig1}). We confine
ourselves with two simple cases of integer nuclear spins, namely $I=1,2$. Then
the overlap integral between the left- and right ground states can be expressed
as
\begin{align}
\eta &  =\cos(\phi),\hspace{2mm}I=1;\nonumber\\
\eta &  =\left\vert \cos^{2}(\phi)-\sin^{2}(\phi)/2\right\vert ,\hspace
{2mm}I=2. \label{eq:spinoverlap_quadr_Bneg_integ}
\end{align}
We will pay most attention to these two non-degenerate cases because they are very 
convenient for the investigation of the effect of an external magnetic field
on the overlap integral (see. Sect. \ref{sect:h}.)

The value of the rotational angle $\phi$ in glasses is unknown. We expect that
it can be estimated by extended molecular dynamics simulations 
\cite{Heuer}. On the other hand, in the orientational glass (KBr)$_{1-x}$
(KCN)$_{x}$ (see Ref. \cite{KBr}) the $CN$ group rotates between different equilibrium
positions by angle $\phi=\cos^{-1}(1/3)$ (see Ref. \cite{Hunklinger}). We assume that in
glasses the rotational angle $\phi$ is the same and neglect its fluctuations due
to a structural disorder. Then, for spin $I=1,$ one can estimate the overlap
integral Eq. (\ref{eq:spinoverlap_quadr_Bneg_integ}) as
\begin{equation}
\eta\approx \cos(\phi) \approx0.33. \label{eq:max_K}
\end{equation}
When the TLS contains $n$ atoms tunneling simultaneously, the characteristic
overlap integrals becomes
\begin{equation}
\eta_{\ast} = \langle l \mid r \rangle\approx\eta^{n}. \label{eq:tot_over}
\end{equation}
To understand and interpret experimental data, we assume that the
total overlap integral is small
\begin{equation}
\eta_{\ast} \ll 1. \label{eq:tot_over1}
\end{equation}
This assumption is justified by the exponential decrease of the
overlap integral with the number of atoms $n$ participating in a single
tunneling system (TLS). 

The case of a half-integer spin is more complicated because of the Kramers
degeneracy. The quadrupole spin Hamiltonian of a tunneling atom $i$ has two
orthogonal ground states $|il+>$, $|il->$ in the left well and two orthogonal ground states $|ir+>$,
$|ir->$ in the right well. As a result, the ground state of $n$ tunneling atoms
is $2^{n}$-fold degenerate. In order to reduce the problem to a single pair of
levels (one in the left well and the other in the right one), we divide the
degenerate ground state into $2^{n}$ pairs of left $|il>$ and right $|ir>$
ground states coupled only with each other. In fact, tunneling amplitude between
left and right states of a tunneling atom $i$ is directly proportional to the
overlap integral of the two states involved $<il|ir>$. Let us introduce two
superposed states in the left well as follows
\begin{align}
\mid l1\rangle &  =\cos(\alpha)\mid il+\rangle+\sin(\alpha)\mid il-\rangle,
\nonumber\\
\mid l2\rangle &  =-\sin(\alpha)\mid il+\rangle+\cos(\alpha)\mid il-\rangle.
\label{eq:new_bas}
\end{align} 
The angle $\alpha$ is defined by the condition of orthogonality
\begin{align}
0  &  =\langle ir+\mid l2\rangle=\langle ir-\mid l1\rangle, \nonumber\\
&  =-\sin(\alpha)\langle ir+\mid il+\rangle+\cos(\alpha)\langle ir+\mid
il-\rangle, \nonumber\\
&  =\cos(\alpha)\langle ir-\mid il+\rangle+\sin(\alpha)\langle ir-\mid
il-\rangle. \label{eq:orth_single_spin}
\end{align}
This equation can be solved for $\alpha$ if the equation determinant is equal to zero, i. e. 
\begin{equation}
\langle il+ \mid ir+ \rangle\langle il- \mid ir+ \rangle+ \langle il+ \mid ir-
\rangle\langle il- \mid ir- \rangle= 0. \label{eq:solubil}
\end{equation}
This condition is satisfied when pairs of eigenstates $|ir+>$, $|ir->$ and
$|il+>$, $|il->$ are orthogonal to each other. One can verify the validity of
Eq. (\ref{eq:solubil}) by projecting the left vectors $|il+>$, $|il->$ onto
the right vector subspace ($|ir+>$, $|ir->$). Choosing the basis for the left
well from the pairs of states Eq. (\ref{eq:new_bas}), we can construct $2^{n}$
possible states of nuclear spins from the products of different left states
(two choices for each atom). The basis states in the right well can be constructed
similarly. Then each of the $2^{n}$ left basis states possesses a non-zero
overlap integral with the only single single state from the $2^{n}$ right basis states.

Thus, the degenerate states form groups of pairs of states connected by
tunneling, while the states belonging to different pairs are uncoupled. Since
all $2^{n}$ pairs possess identical properties with respect to tunneling, one
can treat them as independent pairs of states. Then, our considerations become
similar to those for integer nuclear spins with degenerate ground states of nuclear quadrupoles.

\section{Resonant Susceptibility}

\label{eps}

The main attention of this work is paid to the TLS resonant dielectric
susceptibility $\varepsilon_{res}$ which shows the large deviation from the
conventional tunneling model for $T\leq 5$mK. We begin from reviewing the
nature of the contribution of tunneling systems to the dielectric constant.
Although the resonant dielectric susceptibility has the well established
behavior (for reviews, see \cite{Hunklinger,Phillips_review} and references
therein), our analysis will be helpful for understanding the effect of the
nuclear quadrupole interaction. It is also useful for readers
not too familiar with the TLS dielectric response.

Consider a single TLS polarization due to the external electric field
$\mathbf{F}$. We suppose that tunneling atoms possess a nonzero charge. Then
tunneling between the right and left wells changes the dipole moment of
TLS. The TLS dipole moment operator can thus be expressed in the terms of
pseudospin (see Eq. (\ref{eq:tls_Ham}))
\begin{equation}
\widehat{{\bm\mu}}={\bm\mu}s^{z}. \label{eq:dip_mom_opr}
\end{equation}
Here ${\bm\mu}$ is a dipole moment of a tunneling system. The interaction
of the external field $\mathbf{F}$ with a TLS can be written as
\begin{equation}
\widehat{V}=-\mathbf{F}{\bm\mu}s^{z}. \label{eq:TLS_field_int}
\end{equation}
In general, there can be a contribution to the dipole moment proportional to
the off-diagonal operator $s^{x}$. This results from a change of the barrier
due to electric field. Employing the experimental data \cite{Hunklinger} and
theoretical estimates \cite{B91}, we can argue that such term leads to much
smaller effect than that from the \textquotedblleft diagonal\textquotedblright
term Eq. (\ref{eq:dip_mom_opr}). Therefore we neglect it.

The external field effect application can be taken into account by introducing
the field-dependent asymmetry energy
\begin{equation}
\Delta\left(  \mathbf{F}\right)  =\Delta+\mathbf{F}\widehat{{\bm\mu}}.
\label{eq:TLS_field_int_asym}
\end{equation}
Thus, the energies $E_{\alpha},$ $\alpha=1,2,...Z,$ of all $Z=2\cdot
(2I+1)^{n}$ eigenstates of a tunneling system become dependent on the external
field. Remember that $Z=2$ in the absence of the nuclear quadrupole interaction.

We are interested in a \textit{linear} response of a TLS ensemble, i.e.,
response to an infinitesimal electric field. Then the dipole moment of the
eigenstate $\alpha$ can be expressed as
\begin{equation}
{\bm\mu}_{\alpha} = -\frac{\partial E_{\alpha}}{\partial\mathbf{F}} = -{\bm\mu
}\frac{\partial E_{\alpha}}{\partial\Delta}. \label{eq:mu_alpha}
\end{equation}
The total TLS dipole moment can be expressed as a sum of contributions of all
the $Z$ eigenstates $\alpha$, weighed by the Gibbs population factors
$P_{\alpha}$
\begin{equation}
\overline{{\bm\mu}}=\sum_{\alpha=1}^{Z}{\bm\mu}P_{\alpha}\frac{\partial
E_{\alpha}}{\partial\Delta_{i}}. \label{eq:mu_tot}
\end{equation}
The population factor $P_{\alpha}$ is given by the equilibrium distribution
for \textit{unperturbed} TLS's
\begin{equation}
P_{\alpha} \left(  \Delta_{0}, \Delta; T \right)  = \frac{\exp \left(
-\frac{E_{\alpha}}{T} \right)  }{\sum_{\gamma= 1}^{Z} \exp \left(
-\frac{E_{\gamma}}{T} \right)  }. \label{eq:populat}
\end{equation}
Finally, the susceptibility of a given TLS is determined by
\begin{equation}
\chi_{ab}=\frac{\partial\overline{{\mu}_{a}}}{\partial F_{b}}=\sum_{\alpha
=1}^{Z}\left(  -\mu_{a}\mu_{b}P_{\alpha}\frac{\partial^{2}E_{\alpha}
}{\partial\Delta^{2}}+\mu_{a}\frac{\partial P_{\alpha}}{\partial F_{b}}
\frac{\partial E_{\alpha}}{\partial\Delta}\right). \label{eq:chi_1}
\end{equation}
Here the indexes $a$ and $b$ denote the Cartesian coordinates of the
corresponding vectors and tensors. The first term is associated with the
\textit{adiabatic }excitation of the TLS due to the field-induced change in
its eigenenergies. This contribution reaches its maximum when TLS has an
asymmetry $\Delta$ smaller than its tunneling amplitude $\Delta_{0}$. It is
called a \textit{resonant} contribution \cite{Hunklinger}. The remaining term
is associated with the change in populations of TLS energy levels induced
by the external electric field. Such changes take place by the TLS transitions between
different quantum states. Therefore, this contribution corresponds to the
relaxational term \cite{Hunklinger}.

For temperatures $T<50$mK and an external alternating field $\mathbf{F}$ with
frequency $\nu>100Hz,$ the relaxation contribution is negligibly small
\cite{Hunklinger,DDO1,DDO2} because a relaxation rate of TLS populations
becomes much smaller than the field oscillation rate $\nu$. Therefore, the
relevant TLS's cannot adjust their thermal populations to the rapidly changing
field. In the range of interest $T\sim5$mK we can neglect the relaxational
contribution and restrict our consideration to the first term on the
right-hand side of Eq. (\ref{eq:chi_1}).

The contribution of a single TLS should be summed over all TLS's belonging to
the system. This is equivalent to averaging the susceptibility $\chi_{ab}$ in
Eq. (\ref{eq:chi_1}) over energies, tunneling amplitudes, dipole moments, and
other possible relevant parameters. Averaging over the directions and absolute
values of TLS dipole moments is straightforward and we can rewrite the
resonant TLS contribution in the form
\begin{align}
\chi_{ab}^{res}  
=\delta_{ab}\chi, \ \nonumber\\
\chi =\frac{\mu_{0}^{2}}{3}\left\langle \sum_{\alpha=1}^{Z}P_{\alpha
}\left(  \Delta_{0},\Delta;T\right)  \chi_{\alpha}\left(  \Delta,\Delta
_{0}\right)  \right\rangle,\nonumber\\
\chi_{\alpha}\left(  \Delta,\Delta_{0}\right)   
=-\frac{\partial^{2}E_{\alpha}}{\partial\Delta^{2}}. \label{eq:chi_2}%
\end{align}
Here $\mu_{0}^{2}$ is the average square of the TLS dipole moment. Also, we
have ignored the correlations between TLS dipole moments with the other
tunneling parameters. This is justified by the available experimental data
\cite{Hunklinger}. The average $<...>$ implies the integration of the single TLS
response over $d\Delta d\Delta_{0}/\Delta_{0}$ in accordance with the
postulated TLS distribution Eq. (\ref{Eq:TLSmain}).

Let us first consider the behavior of the dielectric constant in the
zero-temperature limit. In this case it suffices to take into account the
ground state only in Eq. (\ref{eq:chi_1}), calculating the resonant dielectric constant
\begin{equation}
\chi= -\frac{P\mu_{0}^{2}}{3} \int_{0}^{W} \frac{d \Delta_{0}}{\Delta_{0}}
\int_{-\infty}^{\infty} \frac{\partial^{2}E_{g}}{\partial\Delta^{2}} d
\Delta. \label{eq:zeroT}
\end{equation}
The upper integration limit $W$ represents the maximum tunneling amplitude, while the
integration limits for an asymmetry parameter $\Delta$ are set to $\pm\infty$
because its absolute value can be much larger than the tunneling splitting
$\Delta_{0}.$ The integral over asymmetries can be evaluated as
\begin{equation}
\left.  -\frac{\partial E_{g}\left(  \Delta_{0},\Delta\right)  }
{\partial\Delta}\right\vert_{\Delta=-\infty}^{\Delta=\infty}.
\label{eq:one_step_integr}
\end{equation}
For a large asymmetry $|\Delta|\gg\Delta_{0},$ the ground state is determined by
the minimum energy state of a particle in the left $\Delta>0$ (see Fig.
\ref{fig1}) or the right well $\Delta<0$. So, for large $\Delta$ the
ground-state energy behaves as $E_{g}\approx-\mid\Delta\mid/2$, while the
tunneling amplitude and nuclear interaction give rise only to the small
correction. Then the expression in Eq. (\ref{eq:one_step_integr}) equals
unity. Therefore, Eq. (\ref{eq:zeroT}) results in the divergent integral%
\begin{equation}
\chi\approx\frac{P \mu_{0}^{2}}{3} \int_{0}^{W} \frac{d \Delta_{0}}{\Delta
_{0}}. \label{eq:ans_zero_T}
\end{equation}
It is remarkable that this result does not depend on whether or not the
quadrupole interaction exists. So, the question remains: what happens at
finite temperature?

The answer is clear for TLS's without a nuclear quadrupole interaction. In this
case each TLS has only the two states, i.e., the ground state $g$ and the
excited state $e$ with energies $E_{g,e}=\mp(1/2)\sqrt{\Delta^{2}+\Delta
_{0}^{2}}$ (Eq. (\ref{eq:hamtls}), cf. Ref. \cite{Hunklinger}), respectively. The susceptibilities of
excited and ground states differ by the sign since
\begin{equation}
\frac{\partial^{2}E_{e}}{\partial\Delta^{2}} = -\frac{\partial^{2}E_{g}%
}{\partial\Delta^{2}} = -\frac{1}{2} \frac{\Delta_{0}^{2}}{ \left(  \Delta^{2}
+ \Delta_{0}^{2} \right)^{3/2}}. \label{burin}
\end{equation}
The response comes mainly from resonant TLS's having $|\Delta|\leq
\Delta_{0}.$ Therefore, both levels of a TLS with $\Delta_{0}<T$ are
approximately equally populated. For this reason, the contribution of the
excited state to the dielectric constant cannot be neglected. Moreover, the
contributions from the excited and ground state nearly cancel each other if
$\Delta_{0}<T$.

Calculating the finite temperature resonant susceptibility, one should
consider only TLS ground states and cut-off the integral Eq.
(\ref{eq:zeroT}) at the lower limit given by $\Delta_{0}\sim T$. This leads to the
well-known logarithmic temperature dependence
\begin{equation}
\chi= \frac{P\mu_{0}^{2}}{3} \int_{T}^{W} \frac{d \Delta_{0}}{\Delta_{0}}
\int_{-\infty}^{\infty} d \Delta\frac{\partial^{2}E_{g}}{\partial\Delta^{2}} =
\frac{P \mu_{0}^{2}}{3}~ \ln(W/T). \label{eq:finiteT}
\end{equation}

\begin{figure}
[ptbh]
\begin{center}
\includegraphics[
height=1.4105in,
width=2.5434in
]%
{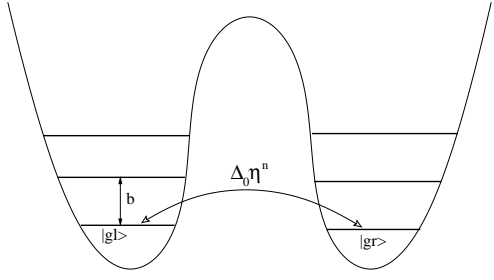}%
\caption{Two level configuration with splitted energy levels.}%
\label{low_T_quadr}%
\end{center}
\end{figure}

This result is valid as long as the temperature exceeds the energy of the
quadrupole interaction $nb$. Next, we discuss the case $T \ll b$. TLS's
with small tunneling amplitudes $\Delta_{0}<nb$ still contribute to the
resonant susceptibility. They can be represented by pairs of lowest nuclear
quadrupole levels in the right and left wells because the higher levels are
separated by the gap $b\gg T$ from these two lowest ones. They are coupled
with each other by the tunneling amplitude $\Delta_{0}$ reduced by the overlap
factor Eq. (\ref{eq:tot_over}) (see Fig. \ref{low_T_quadr}), i.e.,%
\begin{equation}
\Delta_{0 \ast} \approx\Delta_{0} \eta^{n}. \label{eq:ren_tun_ampl}%
\end{equation}
%

These two lowest levels can be treated as a new TLS. Since only TLS's
with $\Delta_{0\ast}>T$ contribute to the permittivity (see Eq.
\ref{eq:finiteT}), this defines the renormalized lower cut-off
\begin{equation}
\Delta_{0l} \sim T \eta^{-n}. \label{eq:newcutoff}
\end{equation}
Substituting this cut-off into integral Eq. (\ref{eq:finiteT}) yields in the limit $T\rightarrow 0$
\begin{equation}
\chi= \frac{P \mu_{0}^{2}}{3} \int_{T\eta^{-n}}^{W} \frac{d \Delta_{0}}%
{\Delta_{0}} \int_{-\infty}^{\infty} d \Delta\frac{\partial^{2}E_{g}}%
{\partial\Delta^{2}} = \frac{P \mu_{0}^{2}}{3} \left(  \ln(W/T) - n\ln(1/\eta)
\right). \label{eq:diel_quadr_ans_2}%
\end{equation}
Thus, due to the quadrupole interaction, we predict qualitatively a noticeable
reduction of the TLS contribution to the dielectric constant at low
temperatures. We believe that this reduction can explain the plateau in the
temperature dependence of the dielectric constant.

It follows from the above analysis that in the whole energy interval
\begin{equation}
\Delta_{0\ast} = \left\{
\begin{array}
[c]{rl}
\Delta_{0}, & \Delta_{0} \gg nb,\\
\Delta_{0}\eta^{n}, & \Delta_{0} \ll nb.
\end{array}
\right.
\label{renorm1}
\end{equation}
This renormalization results in a gap in the distribution of the effective
tunneling amplitude $\Delta_{0\ast}$ of tunneling systems
\begin{equation}
P\left(  \Delta_{0\ast} \right)  = \left\{
\begin{array}
[c]{rll}%
\frac{P}{\Delta_{0\ast}}, &  & \Delta_{0\ast}\gg nb\\
0, & \ nb\eta^{n}\ll & \Delta_{0\ast}\ll nb\\
\frac{P}{\Delta_{0\ast}}, &  & \Delta_{0\ast}\ll nb\eta^{n}.
\end{array}
\right.   \label{renorm2}
\end{equation}
Using the latter result to estimate the dielectric constant temperature
dependence in the expression similar to Eq. ( \ref{eq:diel_quadr_ans_2}), one
can obtain the plateau in the temperature dependence of the dielectric
constant within the range $\ nb\eta^{n}<T<nb$. At $T>nb$ one should use the 
standard tunneling model result Eq. (\ref{eq:finiteT}), while at $T<nb\eta^{n}$ the resonant dielectric constant obeys Eq. (\ref{eq:diel_quadr_ans_2}). 
Thus, our qualitative arguments
can explain the behavior observed experimentally.

In the following part of this paper we investigate the two regimes of Eqs.
(\ref{eq:finiteT}), (\ref{eq:diel_quadr_ans_2}) and
the crossover between them with the higher accuracy.

\section{Perturbation theory approach}

\label{pert_theory}

The renormalization of a tunneling amplitude $\Delta_{0}$ affects the
thermal and kinetic properties of the TLS ensemble and provides a minimum
energy splitting of TLS's having zero asymmetry $\Delta=0.$ In particular, the
resonant susceptibility is determined by similar resonant TLS's with small
asymmetry $|\Delta|<\Delta_{0}.$ Therefore, we will study the most relevant case, i.e., $\Delta=0$. The
resonant permittivity we are interested in here is determined by the ground
state of the tunneling systems with the energy difference between the ground
and first excited states larger than the thermal energy. In this case the
contribution of higher excited states is insignificant because due to
an exponentially small probability of their occupation. A positive contribution
of a ground state to the dielectric constant reflects the general fact that
the ground state minimizes the energy of the system. Thus, the external
electrical field aligns the ground state dipole moment along the field
direction. The susceptibility of excited states can be negative like in the
case of Eq. (\ref{burin}) and also in the case of TLS's is affected by the
nuclear quadrupole interaction as we will show below. Therefore, when the
temperature becomes comparable or larger than energy splitting between 
ground and excited states, TLS susceptibility becomes negligible.
Thus, a structure of the ground state is only important for the resonant
susceptibility of the system.

Next, we consider the effect of the quadrupole splitting on the ground state
of the tunneling system. In the case of vanishing asymmetry energy $\Delta=0$
the Hamiltonian of the tunneling system can be represented in the form
\begin{equation}
H = \frac{\widehat{H}_{r} + \widehat{H}_{l}}{2} + s^{z} \left(\widehat
{H}_{r} - \widehat{H}_{l} \right)  - \Delta_{0}s^{x}.
\label{eq:reson_TLS_int}
\end{equation}
The total wave function of a tunneling particle is a product of the coordinate
(pseudospin) wave function and the nuclear one.

If the tunneling amplitude $\Delta_{0}$ is large, the ground state is described by the wave
function $\left\vert g_{hyb}\right\rangle $ whose coordinate part is symmetric
and the tunneling particles are shared equally between the two wells. In this
case one has $<s^{x}>\approx1/2$, $<s^{z}>\approx0$ and the energy of the
nuclear quadrupoles is given by the \textquotedblleft mean\textquotedblright\ Hamiltonian
\begin{equation}
\frac{\widehat{H}_{r} + \widehat{H}_{l}}{2}. \label{eq:quadr_int_str_tunn}
\end{equation}
This regime is called the hybridized one.

In the opposite limit of the small tunneling amplitude, i.e., in the localized
regime, one can neglect the tunneling term $s^{x}$ in Eq.
(\ref{eq:reson_TLS_int}) and the energy minimum corresponds to either the
ground state of the Hamiltonian $\widehat{H}_{l}$ or to the the ground state
of the Hamiltonian $\widehat{H}_{r}$. Then, the influence of the tunneling
$s^{x}$ term on the ground state is insignificant due to the strong reduction in the
effective tunneling amplitude (see Eq. (\ref{renorm1})) because of the small factor
$\eta^{n}.$ Let us find the crossover between the hybridized and 
localized regimes.

In the hybridized regime the ground state energy is
\begin{equation}
-\frac{\Delta_{0}}{2}+\frac{1}{2}\left\langle g_{hyb}\left\vert \widehat
{H}_{r}+\widehat{H}_{l}\right\vert g_{hyb}\right\rangle, \label{eq:gs}
\end{equation}
while in the localized regime it is
\begin{equation}
\left<  g_{r} \left|  \widehat{H}_{r} \right|  g_{r} \right>  = \left<  g_{l}
\left|  \widehat{H}_{l} \right|  g_{l} \right>. \label{eq:glr}
\end{equation}
Let us introduce a parameter $\lambda_{\ast}$ describing the ground-state
quadrupole energy difference between the two limiting regimes. It is the
reorganization energy corresponding to the transition from the hybridized
state to the localized one
\begin{equation}
\lambda_{\ast} = -\left<  g_{l} \left|  \widehat{H}_{l} \right|  g_{l}
\right>  + \left<  g_{hyb} \left|  \frac{\widehat{H}_{r} + \widehat{H}_{l}}{2}
\right|  g_{hyb} \right>. \label{starlambda}
\end{equation}
Comparing Eqs. (\ref{eq:gs}) and (\ref{eq:glr}), one finds that the
hybridization regime is realized if
\begin{equation}
\Delta_{0} > 2\lambda_{\ast}, \label{eq:delloc_case}
\end{equation}
while in the localized regime this inequality changes its sign, i.e.,
\begin{equation}
\Delta_{0} < 2\lambda_{\ast}. \label{eq:loc_case}
\end{equation}

\begin{figure}
[ptbh]
\begin{center}
\includegraphics[
height=1.8801in,
width=2.5858in
]%
{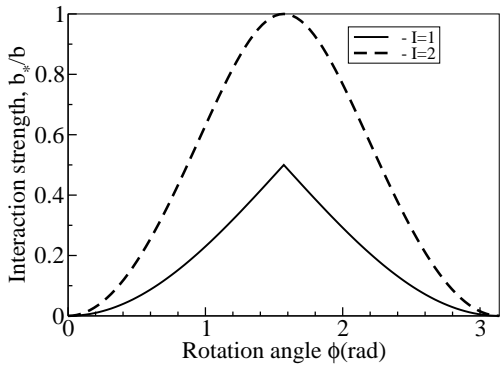}
\caption{Effective interaction energy $b_{\ast}$ of nuclear spins vs. angle
between the nuclear quantization axis in two wells.}%
\label{fig:b_st}%
\end{center}
\end{figure}

Let $n$ be the number of atoms of a TLS experiencing nuclear quadrupole
interaction. The parameter
\begin{equation}
b_{\ast}=\lambda_{\ast}/n \label{b_star}
\end{equation}
represents the reorganization energy per a tunneling atom. Let us estimate the
parameter $b_{\ast}$ for the case when the quadrupole interaction is described
by Eq. (\ref{eq:ax_sym_mod}). One has
\begin{align}
\widehat{H}_{l}  &  = b \left(  I_{x}^{2} - I(I+1)/3 \right),
\label{b*1}\\
\widehat{H}_{r}  &  = b \left(  \left(  I_{x} \cos\phi+ I_{y} \sin\phi\right)
^{2} - I(I+1)/3 \right). \label{b*2}
\end{align}
Then one can calculate the parameter $b_{\ast}$ by using Eqs. (\ref{starlambda}%
), (\ref{b*1}), (\ref{b*2}). To be more specific, consider the case $b>0$ and
integer spin $I=1,2$. The results of the numerical calculation of a single atom parameter $b_{*}$ in the model Eqs. (\ref{b*1}), (\ref{b*2}) for different rotation (mismatch) angles are represented in
Fig. \ref{fig:b_st}. One can see that the reorganization energy is always comparable with the nuclear quadrupole interaction energy.
\bigskip

In the hybridization regime Eq. (\ref{eq:delloc_case}) the quadrupole interaction
is a weak perturbation. Then, the multiplet structure proves to be
insignificant when one calculates the contributions of these tunneling systems
to the susceptibility. In other words, the latter can be calculated within the
standard TLS approach.

However, in the localized regime Eq. (\ref{eq:loc_case}) a multiplet structure
is the decisive feature. In this regime a renormalization of the tunneling
amplitude (\ref{renorm1}) becomes important. If a tunneling amplitude
$\Delta_{0}$ is small enough, the ground state will be a superposition of the
ground states in the isolated left and right wells (see Fig. \ref{low_T_quadr}%
). Let us estimate the upper limit of $\Delta_{0}$ below which the two-level
approximation for the ground state is still valid.

The nuclear-spin ground state in each well can be treated as non-degenerate
(see. Sec. \ref{TLS_ground_state}). One can approximately construct the ground
state and the lowest excited state of the tunneling system as a superposition
of the unperturbed ground states in the two wells. The contrubution of higher
excited states are neglected. This is justified by calculating their
contribution to a lowest-order perturbation theory. We estimate the correction factor
$c$ for the ground state amplitude as 
\begin{equation}
c=1-\sum_{i\neq gr}\frac{\Delta_{0}^{2}\left\vert \langle gl\mid
i\rangle\right\vert ^{2}}{\left(  E_{i}-E_{gl}\right)^{2}},
\label{eq:pert_theor1}
\end{equation}
where $i$ labels all states of the right well.

We determine the parameters of the regime for which the second term can be
neglected as follows. The lowest excited states in each well are separated
from the ground state by some characteristic energy $b\sim\hbar\omega_{0}$
where $\omega_{0}$ is a frequency of the nuclear quadrupole resonance. The
next group of states is separated by the energy gap $\sim 2b$. The vast
majority of states have energies exceeding that for the ground state by
the energy $\lambda\simeq nb$. Because of the large statistical weight, these
states provide the main contribution to Eq. (\ref{eq:pert_theor1}). Therefore
we may replace the denominator in Eq. (\ref{eq:pert_theor1}) by $\lambda$.
Then, the sum of the overlap factors $\mid<gl|i>\mid^{2}$ in the numerator can be well
approximated by unity. The requirement that the second term in Eq.
(\ref{eq:pert_theor1}) is small results in the condition
\begin{equation}
\Delta_{0} < 2\lambda. \label{eq:pert_theor2}
\end{equation}
This condition is weaker than Eq. (\ref{eq:loc_case}). Therefore, $c\approx 1$ is
justified when (\ref{eq:loc_case}) is fulfilled and one can ignore the excited states in both wells.

Thus we can specify two domains of system parameters. One is given
by large tunneling amplitudes $\Delta_{0}>2\lambda_{\ast}$, where the nuclear
quadrupole interaction can be ignored. The other is given by small tunneling
amplitudes $\Delta_{0}<2\lambda_{\ast}$. In the latter case we can restrict
our consideration to the pair formed by the two lowest energy states. These
eigenstates are superpositions of the ground states in the left and right
potential wells. A non-perturbative approach developed in the next section,
where nuclear spins are replaced with oscillators, leads to the similar
results only with some minor deviations.

At this point we turn to the analysis of the dielectric constant using the structure of TLS ground states described above. For an improved analysis, it is convenient to write the resonant dielectric
constant Eq. (\ref{eq:chi_1}) in the following form
\begin{align}
\chi=  \frac{\mu_{0}^{2}}{3}\sum_{a}\left\langle P_{a}\left(  \Delta
_{0},\Delta;T\right)  \chi_{a}\left(  \Delta_{0}=0,\Delta\right)
\right\rangle \nonumber\\
+\frac{\mu_{0}^{2}}{3}\sum_{a}\left\langle \left(  P_{a}\left(  \Delta
_{0},\Delta;T\right)  -P_{a}\left(  \Delta_{0},\Delta;0\right)  \right)
\times\left(  \chi_{a}\left(  \Delta_{0},\Delta\right)  -\chi_{a}\left(
\Delta_{0}=0,\Delta\right)  \right)  \right\rangle \nonumber\\
+\frac{\mu_{0}^{2}}{3}\sum_{a}\left\langle P_{a}\left(  \Delta_{0}
,\Delta;0\right)  \left(  \chi_{a}\left(  \Delta_{0},\Delta\right)  -\chi
_{a}\left(  \Delta_{0}=0,\Delta\right)  \right)  \right\rangle.
\label{eq:chi_mod}
\end{align}
The logarithmic temperature dependence, often serving as an evidence for the
TLS effects \cite{Hunklinger}, comes entirely from the first term. This is is
due to logarithmically uniform distribution of the tunneling amplitudes
Eq. (\ref{Eq:TLSmain}). The logarithmic divergence of the first term is
suppressed for TLS's with the energies of the order of the thermal energy
due to the compensation of a positive contribution from the ground state by
negative contributions of excited states (see below). The third contribution
is a temperature-independent constant. We will ignore it as a background
correction to the susceptibility which does not affect its temperature dependence.

In the absence of the nuclear quadrupole interaction the important first and
second terms can be evaluated. One can reexpress the first \textquotedblleft
logarithmic\textquotedblright\ term as
\begin{equation}
\chi_{log}=\frac{P\mu_{0}^{2}}{3}\int_{0}^{W}\frac{d\Delta_{0}}{\Delta_{0}
}\tanh\left(  \Delta_{0}/(2T)\right), \label{eq:log_term_noquadr}
\end{equation}
and the second \textquotedblleft thermal\textquotedblright\ contribution as
\begin{align}
\chi_{T} =  &  \frac{P\mu_{0}^{2}}{3} \int_{0}^{\infty} \frac{d \Delta_{0}}
{\Delta_{0}} \int_{0}^{\infty} d \Delta\left(  \tanh \left(  \frac{\sqrt
{\Delta^{2} + \Delta_{0}^{2}}}{2T} \right)  - 1 \right) \nonumber\\
&  \times\left(  \frac{\Delta_{0}^{2}}{2 \left(  \Delta^{2} + \Delta_{0}^{2}
\right)  ^{3/2}} - \delta(\Delta) \right). \label{eq:therm_term_noquadr}
\end{align}
In the second integral the upper cut-off for the tunneling amplitude is
replaced by $\infty$ because the integrand decreases exponentially at large energies.

The first contribution can be written as
\begin{align}
\chi_{log}  &  = \frac{P\mu_{0}^{2}}{3} \left(  \ln \left(  \frac{W}{2T}
\right)  - I_{1} \right),\nonumber\\
I_{1}  &  \approx\int_{0}^{\infty} \frac{\ln(x)dx}{\cosh^{2}(x)} \approx
-0.82. \label{eq:chi_fin}
\end{align}
The "thermal" contribution is given by the dimensionless integral
\begin{align}
\chi_{T} =  &  \frac{P\mu_{0}^{2}}{3} \int_{0}^{+\infty} \frac{dx}{x} \int_{-\infty
}^{+\infty} dy \left(  \tanh \left(  \sqrt{x^{2}+y^{2}}-1 \right)  \right)
\nonumber\\
&  \times\left(  \frac{1}{2} \frac{x^{2}}{\left(  x^{2} + y^{2} \right)
^{3/2}} - \delta(y) \right). \label{eq:therm_eff}
\end{align}

It vanishes because the upper limit of the integration over $x$ has been
replaced by $\infty$. This can be demonstrated by using, for instance, the
trigonometric substitution $x=\alpha\cos(\beta),y=\alpha\sin(\beta)$ and
evaluating the integral over $\beta$ first \cite{note1}.

Below we calculate the susceptibility of tunneling systems for different
temperature regimes. According to the above analysis (see Eqs.
(\ref{eq:delloc_case}), (\ref{eq:loc_case})) it is convenient to divide all
tunneling systems into two parts depending on either $\Delta_{0}%
>2\lambda_{\ast}$ or $\Delta_{0}<2\lambda_{\ast}$.

In the first case $\Delta_{0}>nb_{\ast}$ the effect of the nuclear quadrupole
interaction is negligible and one can use the standard two-level approximation
in order to estimate the contribution from these TLS's into the susceptibility as%
\begin{equation}
I_{1} = \int_{2\lambda_{\ast}}^{W} \frac{d \Delta_{0}}{\Delta_{0}}
\tanh\left(  \frac{E \left(  \Delta_{0},0 \right)  }{2T} \right).
\label{3}
\end{equation}
Consider temperatures $T>2\lambda_{\ast}.$ With $E\left(  \Delta_{0},0\right)
=\Delta_{0}$ one can estimate this integral as follows (cf. Eq. (\ref{eq:chi_fin}))
\begin{equation}
I_{1} \approx\ln\frac{W}{2T} - \frac{\lambda_{\ast}}{T}+0.82. \label{6}
\end{equation}
The second term on the right-hand side is small in comparison with the first
one and can be omitted. If the temperature is low $T<2\lambda_{\ast}$, the tangent in
Eq. (\ref{3}) equals unity and we have
\begin{equation}
I_{1} = \ln\frac{W}{nb_{\ast}}. \label{5}
\end{equation}
Now consider the second group of tunneling system for which $\Delta
_{0}<\lambda_{\ast}$. For them the multiplet structure becomes important and
we can confine ourselves to ground state levels of the multiplet in each
well. These levels are coupled by the tunneling amplitude $\Delta_{0}\eta^{n}%
$. Then one has
\begin{equation}
I_{2} = \int_{0}^{2\lambda_{\ast}} \frac{d \Delta_{0}}{\Delta_{0}}
\tanh\left(  \frac{\Delta_{0} \eta^{n}}{2T} \right). \label{8}
\end{equation}
Assume that $T<2\lambda_{\ast}\eta^{n}$. Then
\begin{equation}
I_{2} \approx\ln\frac{nb_{\ast}}{T/\eta^{n}}. \label{10}
\end{equation}
This expression should be added to the term $I_{1}$ given by Eq. (\ref{5}), also
giving a contribution the temperature region considered. Thus, the total
integral is given by
\begin{equation}
I_{1} + I_{2} = \ln\frac{W \eta^{n}}{T}. \label{11}
\end{equation}
In the intermediate temperature range $2\lambda_{\ast}>T>2\lambda_{\ast}
\eta^{n}$ Eq. (\ref{8}) is inapplicable strictly speaking. However, we expect
that the contribution to the dielectric permittivity within the range
$2\lambda_{\ast}>\Delta_{0}>2\lambda_{\ast}\eta^{n}$ is small because this
range corresponds to the gap in the distribution of effective tunneling
amplitudes (Eq. (\ref{renorm2})). In addition, the lower limit of the low
temperature dielectric constant Eq. (\ref{11}) coincides with the upper one of
the dielectric constant in the high temperature range Eq. (\ref{6}). Thus, in
the whole temperature region one has
\begin{align}
\varepsilon_{res}  &  \approx\left(  P\mu^{2}/3\right)  \ln(W/T), &
2\lambda_{\ast}  &  <T,\nonumber\\
\varepsilon_{res}  &  \approx\left(  P\mu^{2}/3\right)  \ln(W/2\lambda_{\ast}),
& 2\lambda_{\ast}\eta^{n}  &  <T<2\lambda_{\ast},\nonumber\\
\varepsilon_{res}  &  \approx\left(  P\mu^{2}/3\right)  \ln(W\eta^{n}/T), & T
&  <2\lambda_{\ast}\eta^{n}. \label{eq:appr_ans}
\end{align}
Our perturbation theory analysis is a phenomenological one and cannot be exact
because we do not know the nuclear spin Hamiltonian. However, it is expected
that the main qualitative features of the systems behavior are reproduced. For
a more quantitative and non-perturbative analysis given in the next section,
we replace the nuclear spins with oscillators and investigate the tunneling
amplitude behavior in such a toy model.

\section{Toy model}

\label{eq:oscill}

Since the nuclear spin interaction in the right and left wells is unknown, we
consider instead a toy model in which nuclear spins are replaced by classical
oscillators. This ``bosonization'' approach to nuclear spins is
justified when one deals with the low energy states of many spins. In
particular, it has been used by Prokofev and Stamp to investigate the nuclear spin interaction
effect on the large spin tunneling \cite{ProkStamp}. We consider a symmetric
TLS ($\Delta=0$) characterized by a coherent tunneling amplitude $\Delta_{0}$
and coupled to $n$ oscillators representing the nuclear spins of $n$ tunneling
atoms forming the TLS concerned. All oscillators have the frequency
$\Omega$ and the mass $M$. These $n$ oscillators are linearly coupled to the TLS and
have the shifted equilibrium positions $x_{i}=\pm x_{0}/2$, $i=1,..n$ when
the TLS occupies the right or left wells, respectively. Then the Hamiltonian of the
system can be expressed as
\begin{align}
\widehat{H}=  &  -\Delta_{0}s^{x}+\sum_{i=1}^{n}\left( \frac{p_{i}^{2}}
{2M}+\frac{M\Omega^{2}x_{i}^{2}}{2}\right) \nonumber\\
&  -s^{z}M\Omega^{2}x_{0}\sum_{i=1}^{n}x_{i}. \label{eq:nosc}
\end{align}
The spin values $s^{z}=\pm1/2$ stand for the TLS residing in the right or left
wells, respectively. The models similar to Eq. (\ref{eq:nosc}) have extensively
been studied within the polaron theory (See e. g. Refs. \cite{KaganPr,Leggett1} and refrences therein). Following the
standard approach one can determine an approximate ground state of the problem
by minimizing the ``classical'' part of Eq. (\ref{eq:nosc}).
This excludes the kinetic energy term of the oscillators. The
``classical'' energy of the system can be expressed as
\begin{equation}
E_{cl}=-\frac{1}{2}\sqrt{\Delta_{0}^{2}+\left(  M\Omega^{2}x_{0}\sum_{i=1}
^{n}x_{i}\right)  ^{2}}+\sum_{i=1}^{n}\left(  \frac{M\Omega^{2}x_{i}^{2}}
{2}\right), \label{eq:nosc_cl}
\end{equation}
where the spin-Hamiltonian has been replaced by its ground state energy,
thereby using the relationship $E_{ground}=-\frac{1}{2}\sqrt{\Delta^{2}%
+\Delta_{0}^{2}}$ for the Hamiltonian $-\Delta s^{z}-\Delta_{0}s^{x}$. Oscillators are treated classically. This is justified when their number is
large, i.e., $n\gg1$. We suppose that this regime is applicable here.

The total energy can be minimized with respect to the oscillator coordinates
$x_{i}$. Then, one has for the derivatives of the classical energy Eq.
(\ref{eq:nosc_cl}) with respect to all coordinates
\begin{align}
0  &  = \frac{\partial E_{cl}}{\partial x_{i}} = -\frac{ \left(  M \Omega
^{2}x_{0} \right)  ^{2} X}{2 \sqrt{\Delta_{0}^{2} + \left(  M \Omega^{2}x_{0}X
\right)  ^{2}}} + M \Omega^{2} x_{i},\nonumber\\
X  &  = \sum_{i=1}^{n} x_{i}. \label{eq:E_cl_min}
\end{align}
These equations can be solved analytically by taking into account that at the
energy minimum all equilibrium coordinates $x_{i}$ are identical, that is
$x_{i}=X/n$. The following relations are found
\begin{align}
&  \Delta_{0}>2\lambda_{\ast}:~~X=0,\nonumber\\
&  \Delta_{0}<2\lambda_{\ast}:~~X=\pm\frac{nx_{0}}{2}\sqrt{1-\left(
\frac{\Delta_{0}}{2\lambda_{\ast}}\right)^{2}},\nonumber\\
&  \lambda_{\ast}=nb_{\ast}, \ \ \hspace{3mm}b_{\ast}=M\Omega_{0}^{2}x_{0}%
^{2}/2. \label{eq:polariz}
\end{align}
The quantity $b_{\ast}$ has been introduced to describe the change in the
oscillator energy induced by its interaction with TLS. For spins, this energy
is equivalent to the quadrupole splitting energy $b_{\ast}$ (see Eq.
(\ref{b_star})).

Thus, depending on the relationship between the oscillator energy
$\lambda_{\ast}$ and the tunneling amplitude $\Delta_{0},$ a TLS ground state
can have different structures. When tunneling is stronger than the TLS
interaction with oscillators, i.e., when $\Delta_{0}>2\lambda_{\ast}$, the
tunneling atoms are distributed equally between the two wells and all
oscillators have their energy minimum at $x_{i}=0$. This state is
energetically most favorable because an identical internal structure for the
left and right states minimizes the energy of the system. In other words, the
tunneling of TLS is so fast that the oscillators see it in its average
position in both wells simultaneously. This state is equivalent to the hybridized 
state considered previously in Sec. \ref{pert_theory}. 

In the opposite case of weak tunneling $\Delta_{0}<2\lambda_{\ast}$ the
symmetry between the right and left wells is broken because of a strong
displacement of oscillators. In this case the system has two energy minima
depending on the sign in the definition of the displacement $X$ in Eq.
(\ref{eq:polariz}). For the nuclear quadrupole interactions, these two states
are represented by the nuclear spin configurations minimizing the energy of
TLS's localized either in the right or left wells.

Both the energy minima are still coupled by tunneling, but the
tunneling amplitude is much weaker then in the case of vanishing oscillator
displacements. Consider the non-adiabatic tunneling regime applicable to the
nuclear spins \cite{ProkStamp}. In this case we may express the effective
tunneling amplitude $\Delta_{0\ast}$ as the product of the coherent tunneling
amplitude $\Delta_{0}$ and the overlap integral $<l|r>$ of the left and right
states of the environment, i.e.,
\begin{equation}
\Delta_{0\ast} = \Delta_{0} \cdot\langle l \mid r \rangle.
\label{eq:modif_ampl}
\end{equation}
In order to estimate the overlap integral, one can use the harmonic approach.
We will use it for the wavefunctions in the left and right wells and consider
the case of zero temperature, reasonable if $T<2\lambda_{\ast}$. The domain
$T\approx2\lambda_{\ast}$, where excited states are important, is relatively
narrow at the large number of oscillators $n$ and can be approximately 
ignored due to a logarithmically weak dependence of the
dielectric constant on $T$.

In the harmonic approach one may expand the energy Eq.(\ref{eq:nosc_cl}) near
the local minima given by Eq. (\ref{eq:polariz}) up to the second order in a
coordinate displacement. This expansion can be written as
\begin{align}
E_{cl}(X)\approx &  -\lambda_{\ast}+\frac{\lambda_{\ast}}{2}\left(  1-\left(
\frac{\Delta_{0}}{2\lambda_{\ast}}\right)  ^{2}\right) \nonumber\\
&  +\frac{1}{2}\sum_{i,j}A_{ij}(x_{i}-X/n)(x_{j}-X/n);\nonumber\\
A_{ij}=  &  ~\frac{\partial^{2}E_{cl}}{\partial x_{i}\partial x_{k}}%
=\frac{4b_{\ast}}{x_{0}^{2}}\left(  \delta_{ij}-\frac{1}{n}\left(
\frac{\Delta_{0}}{2\lambda_{\ast}}\right)  ^{2}\right),
\label{eq:harm_exp}
\end{align}
where $\delta_{ij}$ is the Kronecker symbol. The expansion is written near the
potential minimum at the right well, while for the left well the sign of $X$
should be changed to the opposite.

The harmonic part of the Hamiltonian Eq. (\ref{eq:harm_exp}) can be
represented by $n$ independent modes including the symmetric mode
\begin{equation}
u=\frac{1}{\sqrt{n}}\sum_{i=1}^{n}x_{i},~~\Omega_{s}=\Omega_{0}
\sqrt{1-\frac{\Delta_{0}^{2}}{\lambda_{\ast}}}, \label{eq:vibr_mod_sym}%
\end{equation}
and $n-1$ asymmetric, degenerate modes $\alpha$
\begin{equation}
u_{\alpha} = \sum_{i=1}^{n} c_{i}^{\alpha} x_{i},~~~~ \Omega_{a} = \Omega
_{0},~~~~ \sum_{i} c_{i}^{\alpha} = 0,~~~~ \sum_{i} \left|  c_{i}^{\alpha}
\right|^{2} = 1. 
\label{eq:vibr_mod_sym1}
\end{equation}
A factor $1/\sqrt{n}$ is introduced in Eq. (\ref{eq:vibr_mod_sym}) to conserve
the commutation rules between the coordinate and momentum operators. Only
the symmetric mode interacts with the tunneling motion between the right and
left minima, while asymmetric modes remain unchanged and can therefore be ignored.
Therefore the overlap integral in Eq. (\ref{eq:modif_ampl}) is given by the
overlap between two ground-state wavefunctions of the symmetric harmonic modes
with the mass $M$ and the frequency $\Omega_{s}$, and the equilibrium positions shifted by
$\pm|X|/\sqrt{n}$ Eq. (\ref{eq:polariz}) from the origin. Using the Gaussian wavefunctions for oscillator ground
states, one can express this integral in the form
\begin{align}
\langle l\mid r\rangle &  =\exp\left(  -\frac{M\Omega X^{2}}{n\hbar}\right)
\nonumber\\
&  =\exp\left(  -\frac{nM\Omega x_{0}^{2}}{\hbar}\left(  1-\left(  \frac
{\Delta_{0}}{\lambda_{\ast}}\right)  ^{2}\right)^{3/2}\right).
\label{eq:overlap1}
\end{align}
This result can be used to characterize approximately the tunneling of a TLS
consisting of $n$ atoms coupled to the nuclear spins. The overlap integral for
a single oscillator $i$ is
\begin{equation}
\eta=\langle li\mid ri\rangle=\exp\left(-\frac{M\Omega x_{0}^{2}}{\hbar
}\right)  \label{eq:overlap_single}
\end{equation}
in analogy to the single-spin overlap integral introduced before. The
parameter $\lambda_{\ast}$ can be represented by the single atom quadrupole
splitting energy multiplied by the number of atoms in a TLS, i.e., by
\begin{equation}
\lambda_{\ast} \approx nb_{\ast}. \label{eq:spin_params}
\end{equation}
Thus we may use the following approximate relationship between the initial
(coherent) tunneling amplitude $\Delta_{0}$ and the effective tunneling
amplitude $\Delta_{0\ast}$
\begin{equation}
\Delta_{0\ast}\approx\Delta_{0}\exp\left(-n\ln(1/\eta)\left(  1-\left(
\frac{\Delta_{0}}{\lambda_{\ast}}\right)^{2}\right)^{3/2}\right).
\label{eq:midif_tun}
\end{equation}
In the following we will use this relationship in order to describe the
dielectric response of TLS.

The tunneling amplitude distribution modified by Eq. (\ref{eq:midif_tun}) has
a dip (pseudogap) due to sharp decrease in $\Delta_{0\ast}$ when the coupling strength
$\Delta_{0}$ becomes smaller than the effective rearrangement energy
$\lambda_{\ast}=nb_{\ast}$. We can define the modified distribution of TLS's over
$\Delta_{0\ast}$ for $\Delta_{0\ast}<\lambda_{\ast}$ by using
\begin{align}
&  P(\Delta_{0\ast})\propto\int_{0}^{\infty}\frac{d\Delta_{0}}{\Delta_{0}
}\times\delta\left(  \Delta_{0\ast}-\Delta_{0}\exp\left(  -n\ln(\eta)\left(
1-\left(  \frac{\Delta_{0}}{\lambda_{\ast}}\right)  ^{2}\right)
^{3/2}\right)  \right) \label{pet0}\\
&  =\frac{1}{\Delta_{0\ast}}\frac{1}{1+3n\ln(1/\eta)\left(  \frac
{\widetilde{\Delta}_{0}}{\lambda_{\ast}}\right)  ^{2}\sqrt{1-\left(
\frac{\widetilde{\Delta}_{0}}{\lambda_{\ast}}\right)  ^{2}}},
\label{eq:modif_distr}
\end{align}
where the amplitude $\widetilde{\Delta}_{0}$ is an implicit function of
$\Delta_{0\ast}$ defined by Eq. (\ref{eq:midif_tun}). In the case of a large
tunneling amplitude $\Delta_{0\ast}>\lambda_{\ast}$ the distribution is
logarithmically uniform because $\Delta_{0}=\Delta_{0\ast}$ in that regime.

\begin{figure}
[ptbh]
\begin{center}
\includegraphics[
height=1.919in,
width=2.6671in
]
{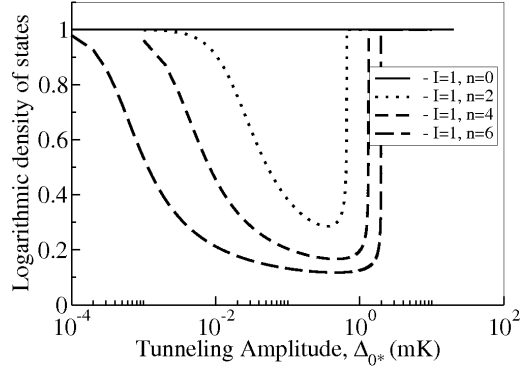}
\caption{The dip in the logarithmic TLS distribution over tunneling amplitudes
due to the nuclear quadrupole interaction. The logarithmic density of
states $P(\Delta_{0}^{\ast})\Delta_{0}^{\ast}$ is shown for $I=1$ and
different numbers $n$ of tunneling atoms. The case $n=0$ corresponds to the
lack of the interaction. The rotation angle is set to $1.23$ rad.}
\label{fig:DOS1}
\end{center}
\end{figure}

\begin{figure}
[ptbhptbh]
\begin{center}
\includegraphics[
height=1.9138in,
width=2.6593in
]
{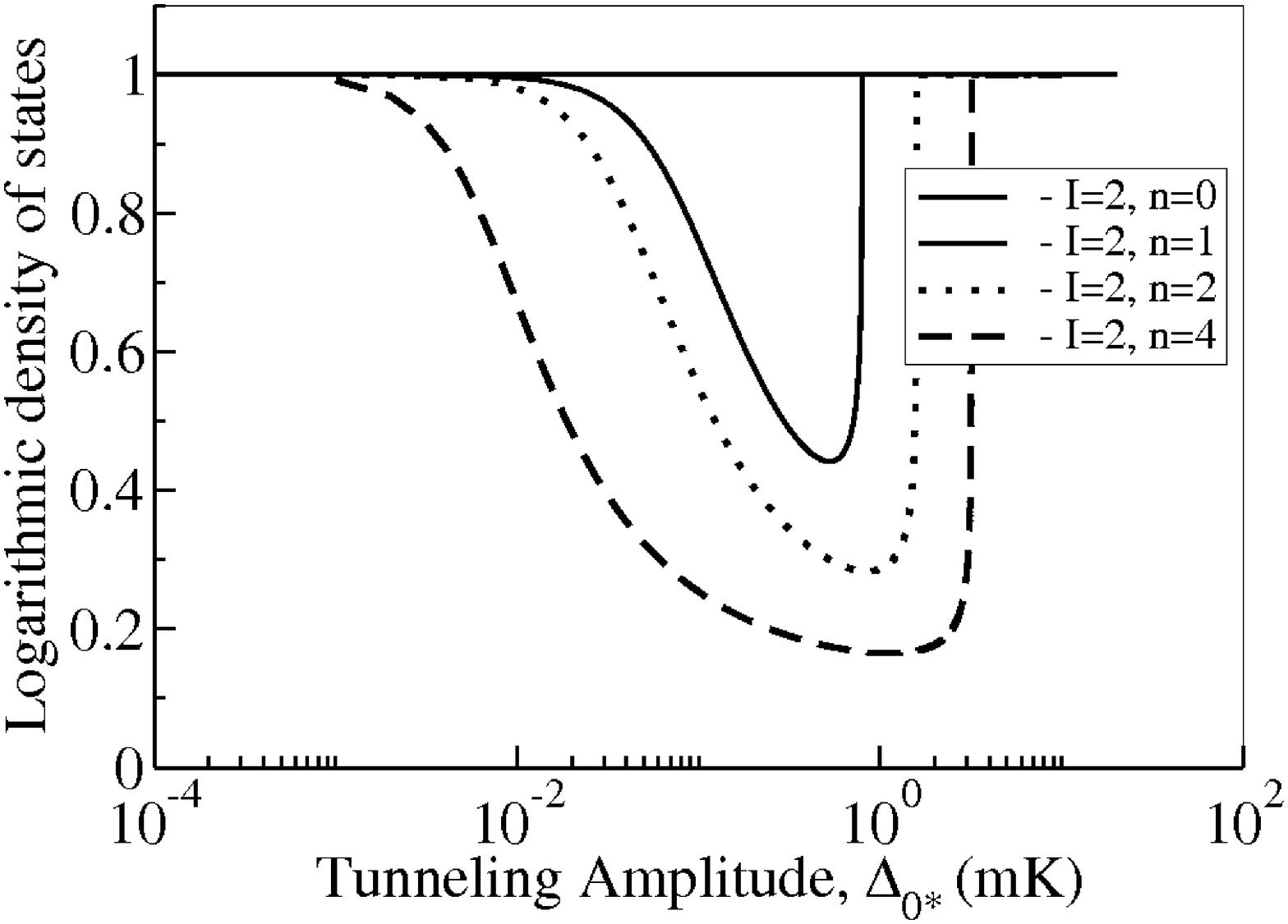}
\caption{The dip in the logarithmic TLS distribution over tunneling amplitude
due to the nuclear quadrupole interactions. The logarithmic density of states
$P(\Delta_{0\ast})\Delta_{0\ast}$ is shown for $I=2$ and different numbers $n$
of tunneling atoms per TLS. }
\label{fig:DOS2}
\end{center}
\end{figure}

For the proof of Eq. (\ref{eq:modif_distr}) we employ
\begin{equation}
\delta\left(  a-\phi\left(  x\right)  \right)  =\frac{1}{\phi^{\prime}\left(
\tilde{x}\right)  }\delta\left(  x-\tilde{x}\right),  \label{pet6}
\end{equation}
where $\tilde{x}$ is the solution of equation $a-\phi\left(  x\right)  =0.$We
identify
\begin{align}
a  &  =\Delta_{0\ast},\ \ x=\Delta_{0}\label{pet2}\\
\phi\left(  x\right)   &  =x\exp\left(  -n\ln(\eta)\left(  1-\left(  \frac
{x}{\lambda_{\ast}}\right)  ^{2}\right)  ^{3/2}\right)  \label{pet3}
\end{align}
and derive
\begin{align}
 \phi^{\prime}\left(  \Delta_{0}\right)  =\exp\left(  -n\ln(\eta)\left(
1-\left(  \frac{\Delta_{0}}{\lambda_{\ast}}\right)  ^{2}\right)
^{3/2}\right) \nonumber\\ \times \left(  1+3n\ln(\eta)\left(  1-\left(  \frac{\Delta_{0}}%
{\lambda_{\ast}}\right)  ^{2}\right)  ^{1/2}\left(  \frac{\Delta_{0}}%
{\lambda_{\ast}}\right)  ^{2}\right)  .\label{pet4}\\
 \nonumber
\end{align}
By substituting Eq. (\ref{eq:midif_tun}) into (\ref{pet4}), we obtain%
\begin{equation}
\phi^{\prime}\left(  \Delta_{0}\right)  =\frac{\Delta_{0\ast}}{\Delta_{0}%
}\left(  1+3n\ln(\eta)\left(  1-\left(  \frac{\Delta_{0}}{\lambda_{\ast}%
}\right)  ^{2}\right)  ^{1/2}\left(  \frac{\Delta_{0}}{\lambda_{\ast}}\right)
^{2}\right). \label{pet5}
\end{equation}
When this expression is put into Eq. (\ref{pet6}) and then into (\ref{pet0}),
we obtain
\[
\int_{0}^{\infty}\frac{d\Delta_{0}}{\Delta_{0}}\times\frac{1}{\frac
{\Delta_{0\ast}}{\Delta_{0}}\left(  1+3n\ln(\eta)\left(  1-\left(  \frac
{\Delta_{0}}{\lambda_{\ast}}\right)  ^{2}\right)  ^{1/2}\left(  \frac
{\Delta_{0}}{\lambda_{\ast}}\right)  ^{2}\right)  _{\Delta_{0}=\tilde{\Delta
}_{0}}}\delta\left(  \Delta_{0}-\tilde{\Delta}_{0}\right)
\]
which coincides with Eq. (\ref{eq:modif_distr}).
The modified distribution of tunneling amplitudes $W(\ln(\Delta_{0\ast
})=P(\Delta_{0\ast})\Delta_{0\ast}$ is shown in Figs. \ref{fig:DOS1},
\ref{fig:DOS2}.


We have chosen a representative value of $\phi=1.23~rad$ for the rotation
angle (see Eq. (\ref{eq:spinoverlap_quadr_Bneg_integ})) in a field of axial
symmetry $bI_{a}^{2}$ and $b>0.$ In accordance with Eq.
(\ref{eq:spinoverlap_quadr_Bneg_integ}) and Fig.\ref{fig:b_st} the effective
interaction parameters are $\eta=1/3$ and $b_{\ast}=0.33$mK for $I=1$, and
$\eta=1/3$ and $b_{\ast}=0.85$mK for $I=2.$ The nuclear quadrupole effect
shows up at an energy which is about three times larger for $I=2$ then for
$I=1$.

One can see from Figs. \ref{fig:DOS1}, \ref{fig:DOS2} that the dip in the
distribution of tunneling amplitudes appears in the domain $(\lambda_{\ast
}\exp(-n\ln(1/\eta)),\lambda_{\ast})$. This becomes sizable for $n=4.$ A similar
behavior is found for the other values of the overlap integral $\eta$. This
dip causes a change of the temperature dependence of the resonant
susceptibility as discussed below.

The resonant dielectric susceptibility of the TLS ensemble can be estimated
from the \textquotedblleft logarithmic\textquotedblright\ integral%

\begin{equation}
\delta\varepsilon\approx\int_{0}^{W}\frac{d\Delta_{0\ast}}{\Delta_{0\ast}%
}P\left(  \Delta_{0\ast}\right)  \tanh\left(  \frac{\Delta_{0\ast}}{2T}\right)
. \label{eq:ets_epsres}
\end{equation}

\begin{figure}
[ptb]
\begin{center}
\includegraphics[
height=1.849in,
width=2.4976in
]%
{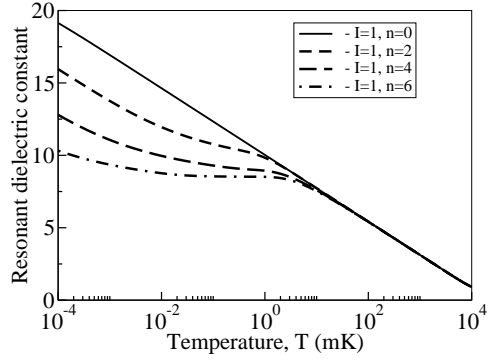}%
\caption{Resonant dielectric constant affected by the nuclear quadrupole
interaction in the case of the nuclear spin $I=1$.}%
\label{fig:eps_S1}%
\end{center}
\end{figure}

\begin{figure}
[ptbptb]
\begin{center}
\includegraphics[
height=1.9614in,
width=2.5676in
]%
{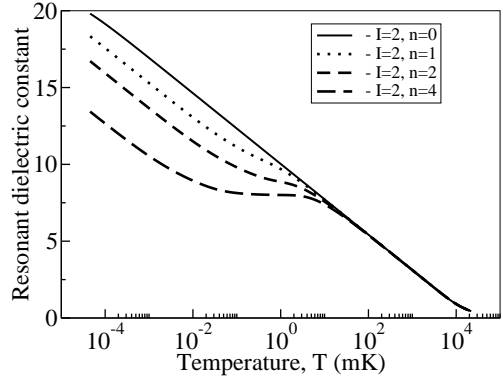}%
\caption{Resonant dielectric constant affected by the nuclear quadrupole
interaction in the case of the nuclear spin $I=2$.}%
\label{fig:eps_S2}%
\end{center}
\end{figure}

By integrating the distribution with the dip we obtain the results shown in
Figs. \ref{fig:eps_S1}, \ref{fig:eps_S2}.%


The quadrupole interaction parameter $b=1$mK and the rotation angle $\phi
\sim1.23$rad are the same as before. One notices a plateau in the dielectric
constant separating two logarithmic temperature dependences characterized by
the same slope at sufficiently large number $n$ of atoms per single TLS. In
the experiment \cite{DDO2,SEH} a plateau is found for $T<5$mK, while the low
temperature edge of the plateau has not been observed because it 
requires much lower temperature than those reached experimentally. A
temperature of $5$mK gives a reasonable estimate of the quadrupole interaction energy.

In all cases the plateau begins at the characteristic maximum temperature
\begin{equation}
T_{max}\simeq\lambda_{\ast}, \label{eq:max_plat_T}
\end{equation}
and extends down to the characteristic minimum temperature
\begin{equation}
T_{min} \simeq\lambda_{\ast} \eta^{n}. \label{eq:min_plat_T}
\end{equation}
For $T<T_{min}$ the behavior of the standard tunneling model is restored and
the dielectric constant shows a logarithmic temperature dependence with the
same slope as at high temperatures $T>T_{max}$. The logarithmic width of the
plateau $\ln\left(  T_{max}/T_{min}\right)  $ is thus directly proportional to
the number of atoms per single TLS.

We expect that in the plateau regime the dielectric constant can be controlled
by the external magnetic field. In fact, the overlap integral of the nuclear
spin states in the left and right well can be affected by the magnetic field.
This effect is discussed in the next section.

\section{Effect of the external magnetic field on the anomalous dielectric
properties at ultra-low temperatures}

\label{sect:h}

The application of the external magnetic field affects the orientation of 
nuclear magnetic moments. When Zeeman splitting becomes comparable with
the nuclear quadrupole interaction, the quantization axes in both potential
wells of the given TLS are aligned with the direction of the magnetic field. Accordingly, the
mismatch of nuclear quadrupole axes will be reduced by the field and the
overlap integral of nuclear quadrupole states in different wells increases and
approaches unity in the high field limit. The effective interaction constant
$b_{\ast}$ vanishes in that case. Thus, high magnetic field reduces the
influence of the nuclear quadrupole interaction and the dielectric constant
behaves like in the standard tunneling model. In this section we estimate the
effect of the external field on the low-temperature resonant dielectric
constant by using the simple oscillator and axially symmetric models for the nuclear quadrupole interaction formulated 
in previous sections.

\begin{figure}
[ptb]
\begin{center}
\includegraphics[
height=1.7884in,
width=2.4803in
]%
{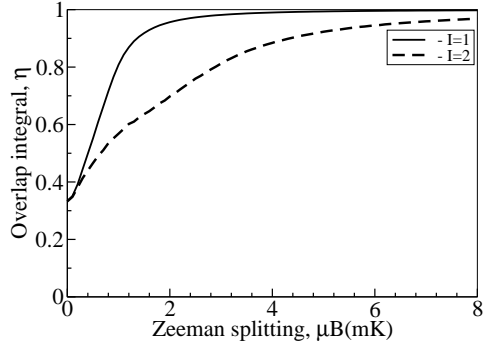}%
\caption{The effect of the magnetic field on the overlap integral. The $1$mK
Zeeman splitting approximately correspond to field $5$T.}%
\label{fig:overlap_vs_h}%
\end{center}
\end{figure}

\begin{figure}
[ptbptb]
\begin{center}
\includegraphics[
width=2.5054in
]%
{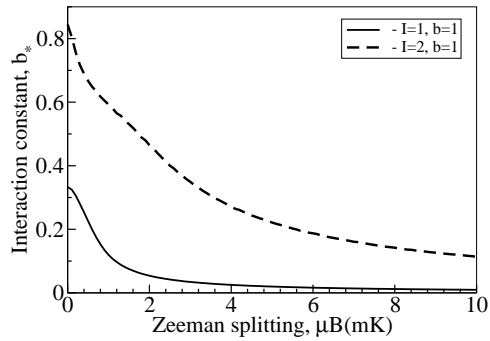}%
\caption{The effect of magnetic field on the effective single-atom interaction
constant $b_{\ast}$}%
\label{fig:b_st_vs_h}%
\end{center}
\end{figure}

Consider the effect of the external magnetic field on parameters
characterizing the mismatch of the nuclear spins in the right and left wells.
The set of parameters necessary to characterize the given TLS includes  
the overlap integral for single nuclear spin $\eta$, the
number $n$ of nuclear spins involved into the tunneling process and the
characteristic nuclear spin interaction energy $b_{\ast}$ (see Eq.
(\ref{eq:midif_tun})). The number $n$ is field-independent, while the overlap integral
$\eta$ and the single atom interaction constant $b_{\ast}$ are subjected to
the changes with the external field. We investigated numerically the change of
the parameters $\eta$ and $b_{\ast}$ in an external field for 
the axially symmetric quadrupole interaction. The nuclear spins are $I=1$ and $I=2$
with the same angle $\cos(\phi)\approx 1/3$ and the nuclear quadrupole
interaction constant $b=1$mK as has been assumed in previous sections. The sign of
the nuclear quadrupole interaction constant $b>0$ is chosen as previously.
Therefore, the ground state has the zero spin projection onto the quantization
axis. The zero-field dielectric constant for these regimes is shown in Figs.
\ref{fig:eps_S1}, \ref{fig:eps_S2}. We have calculated numerically the
dependence on the external magnetic field $B$ of the overlap integral between
the nuclear spin ground states in the left and right wells. Also we have estimated
numerically the dependence of the effective interaction constant on the
magnetic field. The calculations have been made as follows. The quadrupole
Hamiltonian is chosen in the form $b\left(  I^{a}\right)^{2}$ with a the
left-well axis $a=x$ and the right-well axis rotated by the angle $\phi=1.23$rad in the
$x-y$ plane with respect to the $x-$direction. Then the Zeeman term describing
the interaction of the nuclear magnetic moment with magnetic field, is
augmented to the Hamiltonian (\ref{eq:ax_sym_mod}). The magnetic field with a
fixed absolute value $B$ has been generated in the random directions. The
overlap integral of ground states $\eta$ as well as the effective interaction
constant $b_{\ast}$ have been computed and then averaged over $\sim 10^{4}$
realizations of the random field. The results of calculations are shown in
Figs. \ref{fig:overlap_vs_h}, \ref{fig:b_st_vs_h}.%


We express the magnetic field in units of the Zeeman energy splitting (mK). In fact, it is unclear which nuclear spins are most important in the
experiment.  A
reasonable scale we use for estimates is about $1$mK spin splitting in a field
of $B\sim 5$T.

As is clear from Figs. \ref{fig:overlap_vs_h}, \ref{fig:b_st_vs_h}, the
overlap integral increases monotonously with the field and approaches unity
when $\mu B\gg 1$mK. The effective interaction constant $b_{\ast}$ vanishes in
the same limit. Therefore, the application of magnetic field results in the
disappearance of the plateau in the temperature dependence of the dielectric
constant and the standard logarithmic dependence is restored. To examine this
effect, we compute the temperature dependence of the TLS resonant dielectric
constant at various fields. We have used Eq. (\ref{eq:ets_epsres}) with Eq.
(\ref{eq:midif_tun}) for the effective tunneling amplitude and input
parameters $\eta$ and $b_{\ast}$ obtained from our field-dependent
calculations (Figs. \ref{fig:overlap_vs_h}, \ref{fig:b_st_vs_h}).%

\begin{figure}
[ptb]
\begin{center}
\includegraphics[
height=1.9in,
width=2.5668in
]%
{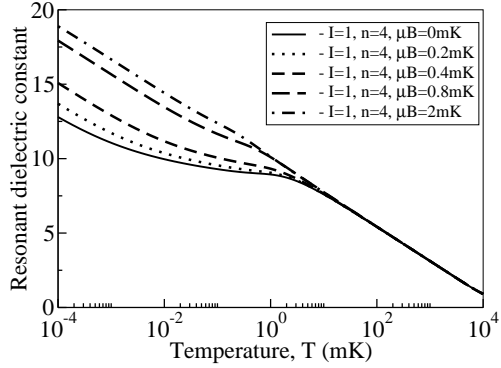}%
\caption{Effect of a magnetic field on the dielectric constant in the case of
spin $I=1$ and the number of atoms per TLS $n=4$.}%
\label{fig:eps_h_S1}%
\end{center}
\end{figure}
\begin{figure}
[ptbptb]
\begin{center}
\includegraphics[
height=1.8135in,
width=2.4491in
]%
{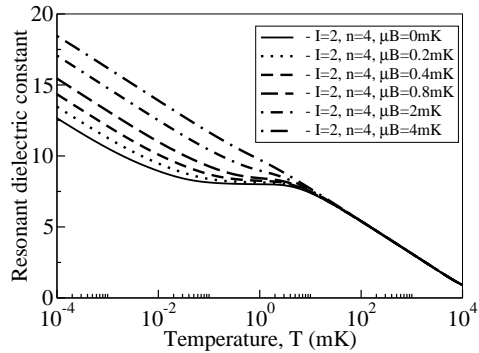}%
\caption{The effect of magnetic field on the dielectric constant in the case
of spin $I=1$ and the number of atoms per TLS $n=4$.}%
\label{fig:eps_h_S2}%
\end{center}
\end{figure}

The field-dependent dielectric
constant is shown in Figs. \ref{fig:eps_h_S1},
\ref{fig:eps_h_S2} for $n=4$ atoms per single TLS and nuclear spins $I=1$ and
$I=2$, respectively. The field effect is similar in both cases. However, 
a stronger field is needed to suppress the nuclear quadrupole
interaction in the case of larger spin $I=2$. 

We can interpret the field dependence as follows. A
relatively small field $\mu B\leq0.2$mK and $B\leq 1$T affects the resonant
dielectric constant only for low temperatures $T\leq 5$mK. At higher
temperatures the field effect can not be viewed beeing a small correction. The
further increase of the field leads to linear reduction in the width of the
plateau from both the high- and low-temperature sides. When the field is
sufficiently high $\mu B\geq1$mK, and $B\geq 5$T, it competes with the nuclear
quadrupole splitting and the plateau narrows by orders of magnitude and
almost vanishes at $\mu B\sim 4$mK. Then the standard tunneling model behavior
is restored. It would be very interesting to perform measurements of the
dielectric constant at very low temperatures, i.e., for $T\leq 5$mK and a strong
magnetic field around $10$T. In this case the plateau in the dielectric
constant should disappear. We hope that our work will stimulate such measurements.

\section{Relationship of the model to real systems}

\label{exp_params}

In this section we address two questions. First question is whether our choice of parameters for the nuclear quadrupole interaction is justified and these parameters can be used to fit the existing experimental data. The second question is related to glasses having no (or negligable) nuclear quadrupole interaction. These glasses (mylar, SiO$_{2}$) should not show any anomalies in the dielectric constant temperature dependence. The experimental data related to these glasses are discussed below. 

According to the previous consideration we expect that the plateau
observed experimentally in the temperature dependence of the dielectric
constant for $T\leq 5$mK is due to the fact that the parameter $nb$ (see Eq.
(\ref{eq:Quadr_int1})) amounts to a value of $5$mK. Experimentally, a
qualitative change in the TLS resonant dielectric susceptibility is seen in
the temperature range of $T\sim5-10$mK (see Refs. \cite{SEH,DDO2,b6}). There is no problem to intepret the saturation behavior using our formalism within the experimental accuracy choosing properly the fitting parameters \cite{prl2005}. 
However the question
arises whether or not the nuclear quadrupole interaction is sufficiently
strong to display itself in this temperature range. We also expect that
glasses lacking the nuclear quadrupole interaction should not show any
deviations from the standard model.

The nuclear quadrupole interaction constant $b$ is defined by Eq.
(\ref{eq:Quadr_int1}). The values of the electric field gradient are close to
each other in different glasses, while the nuclear quadrupole moments $Q$  differ strongly for different elements, leading to the
broad distribution in the observed energies of the nuclear quadrupole
resonance $\hbar\omega_{0}\sim b$ Ref. \cite{H}. The data for the chemical elements
present in glasses for which the low temperature measurements have been made
are summarized in Table 1.\newline%
\bigskip
\begin{tabular}
[c]{|l|l|l|l|l|}\hline
Nucleus & $I$ & $\mu_{N}$ & $Q$(barn) & $\hbar\omega_{0}$(mK)\\\hline
$^{1}$H & $1/2$ & $4.837$ & $0$ & $0$\\\hline
D$=^{2}$H & $1$ & $1.213$ & $0.00286$ & $0.7\cdot10^{-2}$ Ref. \cite{D1}\\\hline
$^{12}$C & $0$ & $0$ & $0$ & $0$\\\hline
$^{16}$O & $0$ & $0$ & $0$ & $0$\\\hline
$^{23}$Na & $3/2$ & $2.863$ & $0.104$ & $0.14$ (Na), $0.4$ (NaF)
Ref. \cite{Na}\\\hline
$^{27}$Al & $5/2$ & $4.309$ & $0.147$ & $0.9$ in Al Ref. \cite{Al}\\\hline
$^{29}$Si & $1/2$ & $-0.962$ & $0$ & $0$\\\hline
$^{29}$K & $3/2$ & $0.505$ & $0.0585$ & -\\\hline
$^{135}$Ba & $3/2$ & $1.82$ & $0.160$ & $0.86$ (BaBiO$_{3}$) Ref. \cite{Ba}\\\hline
\end{tabular}

\bigskip

{\small Table 1.~~ \begin{minipage}[t]{13cm}
Nuclear spin, magnetic moment, quadrupole moment and frequency of nuclear
quadrupole resonance for different chemical elements, possibly participating in
the tunneling.
\end{minipage}}

\vspace{1cm}

As one can see from Table 1, the typical range of nuclear quadrupole
interactions for Na, K, Al, Ba nuclei is around but slightly less than
$1$mK. This interaction should be larger in glassy materials because the
above-mentioned elements are bound with non-metal atoms there. The covalent and
ionic bonds are expected to be stronger than metallic bonds, in particular,
due to the absence of screening. Therefore, their binding energy and,
accordingly, the electric field gradient affecting the nuclear quadrupole
interaction are expected to be larger in these glasses.

\begin{tabular}
[c]{|l|l|l|l|l|}\hline
Glass & Nuclei & $T_{sat}$ (mK)   & Exp. & Refs.\\\hline
Mylar (C$_{10}$H$_{8}$O$_{4}$)$_{n}$& No & $<1$mK & $\varepsilon$ & Ref. \cite{DDO2}\\\hline
$5\%$K-SiO$_{2}$ & K & $4$mK & $\varepsilon$ & Ref. \cite{DDO2}\\\hline
$10\%$K-SiO$_{2}$ & K & $4$mK & $\varepsilon$ & Ref. \cite{DDO2}\\\hline
BK7 \cite{BK7}  & Na & $5$mK & $\varepsilon$ & Ref. \cite{DDO2}\\\hline
SiO$_{x}$ & No & $8$mK & $\varepsilon$ & Ref. \cite{DDO2}\\\hline
BaO-Al$_{2}$O$_{3}$-SiO$_{2}$ & Al,Ba & $\sim 5$mK & $\varepsilon$ &
Ref. \cite{SEH}\\\hline
$a-$SiO$_{2}$ & No & $<2$mK & $v$ & Ref. \cite{Pohl2,MoreRecent}\\\hline
BK7 & Na & $<5$mK & $v$ & Ref. \cite{BK7Ac}\\\hline
\end{tabular}

\bigskip{\small Table 2.~~ \begin{minipage}[t]{13cm}
Saturation temperature below which the dielectric constant
$\varepsilon$ and/or sound velocity $v$ become temperature-independent.
\end{minipage}}

\vspace{1cm}

The experimental data indicating the
strong changes in low temperature dielectric and, probably, acoustic
properties are summarized in Table 2. The saturation in a temperature dependence of the dielectric constant below the
temperature $T_{sat}$ takes place in all materials containing Na, K, Al
or Ba which have relatively high quadrupole moments (see Table 1). The high
saturation temperature observed in BaO-Al$_{2}$O$_{3}$-SiO$_{2}$ is most likely
due to large values of the Ba and Al quadrupole moments compared with the
other elements in Table 1. 

The absence of the low temperature saturation
in the dielectric constant of mylar is in the full agreement with the theory. Indeed, mylar is
an organic polymer composed of C, H and O atoms, for which the most stable isotopes have
vanishing nuclear quadrupole moments. Similarly according to the most recent experimental data 
\cite{Pohl2,MoreRecent} there is no saturation in the temperature ependence of the sound velocity in $a-$SiO$_{2}$ having no nuclear quadrupole interaction. 
A saturation in the low
temperature dependence of the dielectric constant has also been observed in
SiO$_{x}$ which poses a puzzle. One possible explanation is that unpaired
electrons may be present in this material \cite{DNPrivat}. They act like
nuclear quadrupole moments in that aspect.

The problem of interpretation of the acoustic experiment \cite{BK7Ac} is a
knotty one. If the same two-level systems control the acoustic and dielectric
behaviors of the glasses, the saturation should take place below the same
$T_{sat}$ both for the dielectric constant and for the sound velocity
\cite{Hunklinger}. On one hand, there is no saturation in the logarithmic
temperature dependence in $\alpha-$SiO$_{2}$ having no nuclear quadrupoles in
agreement with the theoretical expectations. It is difficult, however, to
understand the absence of any deviations from the standard tunneling model for
the sound velocity measurements in BK7 down to $5$mK Ref. \cite{BK7Ac}. Although
the major contribution to the dielectric constant and the sound velocity can
be due to different subsets of TLS's possessing either larger dipole moments
or stronger elastic coupling to lattice vibrations \cite{DDO5}, there
must be some amount of TLS's containing quadrupole nuclei and contributing to
both dielectric and acoustic properties. These TLS's should lead to an
anomalous acoustic behavior. One possible explanation of the absence of any
saturation effect is that the measurements have been made for $T>5$mK, while
$T_{sat}\approx 5$mK (see Table 2). An extension of acoustic measurements in
BK7 to lower temperatures should help to clarify the puzzling situation.

\section{Discussion and Conclusions}

\label{sect:disc}

In this work we have considered various aspects of the effect of the nuclear
quadrupole interaction on the low temperature properties of glasses. The
significance of this interaction has been pointed out by W\"{u}rger,
Fleischmann and Enss \cite{WFE}. In the present paper this interaction 
has been employed to characterize the
resonant dielectric susceptibility of amorphous solids at ultra-low
temperatures $T\leq 5$mK where major deviations from the predictions of the
tunneling model are observed. The analysis of the typical interaction
parameters shows that the nuclear quadrupole interaction is strong enough to
explain these deviations. To our knowledge the materials having no nuclear quadrupole 
interaction and no unpaired electrons, i. e. the organic polymer mylar \cite{DDO1}, and $\alpha$-SiO$_{2}$ Ref. \cite{Pohl2,MoreRecent} do not show the deviations from the logarithmic 
temperature dependence of a dielectric constant or a sound velocity down to the 
lowest temperatures accessible experimentally around $T\sim 1$mK. These results support our theory. 

Our theory uses the fact that the nuclear quadrupole interaction is generally different in the right and
left wells of a two-well tunneling system. Therefore, it affects the coherent
coupling between the wells. The overall quadrupole interaction effect is
governed by the relative magnitude of two parameters, i.e. $\lambda_{\ast
}=nb_{\ast}$ which is the characteristic nuclear quadrupole interaction strength of TLS's consisting of
$n$ atoms and the tunneling matrix element $\Delta_{0}$ between the left and
right wells. When $\Delta_{0}>\lambda_{\ast}$, the nuclear quadrupole
interaction can be neglected. At smaller tunneling amplitude 
$\Delta_{0}<\lambda_{\ast}$ the effective tunneling coupling  is reduced
exponentially due to the small overlap between different nuclear spin
eigenstates in different wells. This is similar to the polaron effect and we
use this similarity in order to suggest a solvable toy model based on the
replacement of the nuclear spins with oscillators.

The resonant dielectric constant is determined by the contribution of
all resonant TLS's with a characteristic tunneling amplitude of order of energy, 
which varies from the
thermal energy to some characteristic maximum value $T<\Delta_{0}<W$. The
deviations from the standard tunneling model are observed at temperatures
comparable with the nuclear quadrupole interaction, i.e., for $T\simeq
\lambda_{\ast}$. They show up as
a temperature-independent plateau due to the breakdown of coherent tunneling in
the energy range $T<\Delta_{0}<\lambda_{\ast}$, where resonant TLS's do not exist. 
Therefore, TLS's with tunneling amplitudes belonging to this domain do not contribute to the
resonant dielectric constant. Note that the hypothesis of the breakdown of
coherent tunneling below $T \sim 10mK$ has been proposed by Enss and Hunklinger \cite{HE1} assuming,
however, that it is due to the interaction between TLS's. Our work suggests 
the valuable realization of their hypothesys. 

The results agree qualitatively with a number of low-temperature dielectric
measurements made by different groups. To obtain quantitative agreement for
the temperature at which the plateau forms, we have to assume that the
relevant two-level systems consist of, at least, four atoms $\left(
n=4\right)$. This assumption agrees qualitatively with the TLS model based
on the renormalization group theory \cite{Burin_Kagan} and is in line with molecular
dynamics studies of glasses \cite{Heuer}.

At very low temperatures $T<T_{min}$ ($T_{min}<0.1mK$) the plateau in the resonant dielectric constant 
should go over into a logarithmic temperature dependence characterized by the
same slope $d\varepsilon_{res}/d\ln(T)$ as found at high temperatures, i.e.,
for $T>T_{max}\sim \lambda_{\ast}$. Therefore, it is worth
while attempting an experimental observation of this behavior. It is unclear
whether it is possible to perform measurements at such low temperatures at
present. A comparison of the measurements with the theory can be used to
estimate the number $n$ of tunneling atoms involved into single TLS because the
logarithmic width of the plateau $\ln(T_{max}/T_{min})$ is approximately equal $n$.

The nuclear quadrupole interaction should affect the sound velocity since it
enters its resonant part in the same manner as to the resonant dielectric susceptibility.
Although it is clear from available experiments that there is a difference
in the nature of TLS's contributing to dielectric and acoustic properties,
there are observed correlations between the two responses both in the
hole burning and non-equilibrium dielectric measurements
\cite{HH,Hunklinger,DDO1}. We propose to extend acoustic measurements of
glasses possessing atoms with nuclear quadrupole moments to lower temperatures
in order to search deviations in the temperature dependence of a 
sound velocity. The most promising candidates are those materials in which the plateau
in the dielectric constant is seen, including BK7, $10\%$ K-SiO$_{2}$ and
BaO-Al$_{2}$O$_{3}$-SiO$_{2}$ glasses (see Table 2).

We have also analyzed the dependence of the resonant dielectric constant on
the external magnetic field. We show that a high magnetic field $B \sim 5-10$T 
affects nuclear spins stronger than the nuclear
quadrupole interaction and therefore will restore coherent tunneling. This is
because the nuclear spin states in the right and left wells get aligned
parallel to the field. We have performed the model calculations of the dielectric constant in a 
magnetic field by combining numerical simulations for the behavior of a
single spin affected by both a quadrupole interaction and an external field. 
The experimental verification of our theory can be made measuring the dielectric 
constant in the strong external magnetic field, which should eliminate the 
temperature independent plateau and restore the logarithmic temperature dependence.

To our knowledge, most of experimental studies of the magnetic field effect on
the dielectric constant have been performed at relatively high temperatures,
i.e., for $T>10$mK Ref. \cite{SEH}. These studies show certain similarities and
also distinctions compared to our predictions. The major changes in the
dielectric constant are observed at $T\simeq 5$mK. The application of magnetic
field leads to an increase in the dielectric constant in agreement with
expectations. At a certain magnetic field the dielectric constant of the
standard tunneling model is restored. However,the change in the dielectric
constant does not show a monotonous dependence on the magnetic field strength.
A relatively strong effect is observed at very small fields (for $T\simeq 30$mK
and $B\simeq 0.1$T). So, the behavior of the system is more complicated in this
temperature range. Perhaps, the relaxational contribution to the dielectric
constant is still significant there. Then, both the phonon-stimulated relaxation
\cite{wurger1} and the interaction-induced relaxation \cite{PFB} of TLS's
could be affected by an external magnetic field and our analysis would not be
applicable. The experiments at lower temperatures $T\simeq 5$mK can test the
model proposed here and can be used to characterize the internal structure and
nuclear quadrupole interaction for a single TLS.

The work of AB and YS is supported by the Louisiana Board of Regents (Contract
No. LEQSF (2005-08)-RD-A-29 and the LINK Program of the Louisiana EPSCORE 
and TAMS GL fund (account no. 211043) through the Tulane University, College of
Liberal Arts and Science. AB also greatly appreciates the hospitality of Boris
Shklovskii and other members of the William P. Fine Institute of Theoretical
Physics at the School of Physics and Astronomy of the University of Minnesota,
who made it possible for him to stay with the University of Minnesota after
the hurricane disaster in the city of New Orleans. The work of IYP is
supported by the Russian Fund for Basic Researches and the Russian
goal-oriented scientific and technical program "Investigations and
elaborations on priority lines of development of science and technology"
(Contract RI -112/001/526 ).

Authors are also grateful to Siegfried Hunklinger, Robert Pohl, Christian
Enss, Douglas Osheroff, Andreas Fleischmann, Alois Wuerger, Clare Yu, Jeevak
Parpia, Doru Bodea, Andrew Fefferman and Lidia Polukhina for many useful and
stimulating discussions.

In addition, authors acknowledges the organizers and participants of the
International Workshop on Quantum Disordered Systems, Glassy Low-Temperature
Physics and Physics at the Glass Transition (Dresden 2003) for useful discussions.


\begin{thebibliography}{99}                                                                                               %


\bibitem {ahvp}P. W. Anderson, B. I. Halperin, C. M. Varma, Phil. Mag.
\textbf{25}, 1 (1972); W. A. Phillips, J. Low Temp. Phys. \textbf{7}, 351 (1972).

\bibitem {Zeller}R. C. Zeller, R. O. Pohl, Phys. Rev. B \textbf{4}, 2029 (1972).

\bibitem {Hunklinger}S. Hunklinger, A. K. Raychaudchary, Progr. Low Temp.
Phys. \textbf{9}, 267 (1986).

\bibitem {Phillips_review}W. A. Phillips, Rep. Prog. Phys. \textbf{50}, 1567 (1987).

\bibitem {DDO1}D. J. Salvino, S. Rogge, B. Tigner, D. D. Osheroff, Phys. Rev.
Lett. \textbf{73}, 268 (1994).



\bibitem {Barbara}R. Giraud, W. Wernsdorfer, A. M. Tkachuk, D. Mailly, B.
Barbara, Phys. Rev. Lett. \textbf{87} , 057203 (2001).

\bibitem {Stamp}N. V. Prokofev, P. C. E. Stamp, J. Low Temp. Phys.
\textbf{104} , 143 (1996).

\bibitem {Rosenbaum}S. Ghosh, R. Parthasarathy, T. F. Rosenbaum, G. Aeppli,
Science \textbf{296}. 2195 (2002).

\bibitem {Klinger}V. G. Karpov, M. I. Klinger, F. N., Ignat'ev, Sol. State
Commun. \textbf{44}, 333 (1982).

\bibitem {Parshin}V. G. Karpov, D. A. Parshin, JETP \textbf{61}, 1308 (1985).

\bibitem {Leggett}J. J. Freeman, A. C. Anderson, Phys. Rev. B \textbf{34},
5684 (1986); A. J. Leggett, C. C. Yu, Comments Cond. Matter Phys.\textbf{14},
231 (1989).

\bibitem {Burin_Kagan}A. L. Burin, Yu. Kagan, JETP \textbf{82}, 159 (1996);
Phys. Lett. A \textbf{215}, 191 (1996).

\bibitem {b6}A. L. Burin, D. Natelson, D. D. Osheroff, Yu. Kagan, Chapter 5 in
\textquotedblright Tunneling Systems in Amorphous and Crystalline
Solids\textquotedblright, ed. P. Esquinazi, p. 243 ( Springer, 1998).

\bibitem {BurinJLTP}A. L. Burin, J. Low Temp. Phys. \textbf{100}, 309 (1995).

\bibitem {DDO2}S. Rogge, D. Natelson, B. Tigner, D. D. Osheroff, Phys. Rev. B
\textbf{55}, 11256 (1997).

\bibitem {SEH}P. Strehlow, M. Wohlfahrt, A. G. M. Jansen, R. Haueisen, G.
Weiss, C. Enss, S. Hunklinger, Phys. Rev. Lett. \textbf{84}, 1938 (2000).

\bibitem {WFE}A. W\"urger, A. Fleischmann, C. Enss, Phys. Rev. Lett.
\textbf{89}, 237601 (2002).

\bibitem {D1}P. Nagel, A. Fleischmann, S. Hunklinger, C. Enss, Phys. Rev.
Lett. 92, 245511 (2004). 


\bibitem {EnssReview}C. Enss, Physica B \textbf{316}, p. 12 (2002).

\bibitem{Ludwig} 
S. Ludwig, P. Nagel, S. Hunklinger, C. Enss, J. Low Temp. Phys. {\bf 31}, 89 (2002).  

\bibitem {prl2005}A. L. Burin, I. Ya. Polishchuk, P. Fulde, Y. Sereda, Phys.
Rev. Lett. (to appear) (2005).

\bibitem {Heuer}J. Reinisch, A. Heuer, Phys. Rev. B \textbf{70}, 064201
(2004); A. Heuer, R. J. Silbey, Physica B \textbf{220}, 255 (1996).

\bibitem {KBr}J. N. Dobbs, M. C. Foote, A. C. Anderson Phys. Rev. B
\textbf{33}, 4178 (1986).

\bibitem {B91}A. L. Burin, JETP Letters, \textbf{54}, 320 (1991); A. L. Burin,
Fizika Nizkikh Temperatur \textbf{17}, 872 (1991).

\bibitem {note1}This estimation is valid for constant parameter $P$. It can be
shown that the weak dependence of the parameter $P$ on the asymmetry $\Delta$
and the energy splitting $\Delta_{0},$ due to the dipole gap, influences this
conclusion insignificantly, only.

\bibitem {ProkStamp}N. V. Prokof'ev, P. C. E. Stamp, Reports on Progress in
Physics \textbf{63}, 669 (2000).

\bibitem {KaganPr}Yu. Kagan, N. V. Prokofev, Quantum Tunneling Diffusion in
Solids, in ``Quantum Tunneling in Condensed Media", eds. Yu. Kagan, A. J.
Leggett (Elsevier; Amsterdam, 1992) p. 37.

\bibitem {Leggett1}A. J. Leggett, S. Chakravarty, A. T. Dorsey, M. P. A.
Fischer, and W. Zwerger, Rev. Mod. Phys. 59, 1 (1987).

\bibitem {H}The information is taken from the website
"http://www.webelements.com'', based on several sources from literature,
including R. K. Harris in \textit{Encyclopedia of Nuclear Magnetic Resonance},
D. M. Granty and R. K. Harris, (eds.), vol. 5, John Wiley and Sons, Chilester,
UK, 1996; D. R. Lide, (ed.), CRC \textit{Handbook of Chemistry and Physics
1999-2000}: A ready-reference book of chemical and physical data (CRC Handbook
of chemistry and physics, CRC Press, Boka Raton, FL, USA, 79th edition, 1998;
Table of Nuclear Magnetic Dipole and Electric Quadrupole Moments, see
$http://www.nndc.bnl.gov/nndc/stone_{o}ments/nuclear-moments.pdf$.

\bibitem {Na}D. Sundholm, J. Olsen, Phys. Rev. Lett. \textbf{68}, 927 (1992);
C. D. Hollowell, A. J. Hebert, K. Street, Jr., J. Chem. Phys. \textbf{41},
3540 (1964).

\bibitem {Al}N. J. Martin, P. G. H. Sandars, G. K. Woodgate, Proc. R. Soc.
London A \textbf{305}, 139 (1968); J. S. M. Harvey, L. Evans, H. Lew, Can. J.
Phys. \textbf{50}, 1719 (1972).

\bibitem {Ba}M. M. Savosta, V. D. Doroshev, V. A. Borodin, Yu. G. Pashkevich,
V. I. Kamenev, and T. N. Tarasenko, J. Englich, J. Kohout, A. G. Soldatov, S.
N. Barilo, S. V. Shiryaev, Phys. Rev. B \textbf{63}, 184106 (2001).

\bibitem{BK7} 
The chemical structure of BK7 is given by (SiO$_{2}$)$_{0.748}$(B$_{2}$O$_{3}$)$_{0.96}$)(Na$_{2}$O)$_{0.11}$(K$_{2}$O)$_{0.047}$, see Ref. \cite{Ludwig}. 

\bibitem {Pohl2}E. Nazaretski, R. D. Merithew, R. O. Pohl, J. M. Parpia, Phys.
Rev. B \textbf{71}, 144201 (2005).

\bibitem {MoreRecent}A. Fefferman, J. M. Parpia, to appear in the Materials of
the Low Temperature Physics International Conference (Orlando, Florida, USA,
2005); private communication.


\bibitem {BK7Ac}M. Layer, M. Heitz, J. Classen, C. Enss, S. Hunklinger, J. Low
Temp. Phys. \textbf{124}, 419 (2001).

\bibitem {DNPrivat}D. Natelson, Private Communication.



\bibitem {DDO5}D. Natelson, D. Rosenberg, D. D. Osheroff, Phys. Rev. Lett.
\textbf{80}, 4689 (1998).

\bibitem {HE1}C. Enss, S. Hunklinger, Phys. Rev. Lett. \textbf{79}, 2831 (1997).

\bibitem {HH}S. Hunklinger, W. Arnold, in \textit{Physical Acoustics}, ed. by
W. P. Mason, R. N. Thurston (Academic, N. Y. 1976), 12, 155.

\bibitem {wurger1}A. W\"urger, J. Low Temp. Phys. \textbf{137}, 143 (2004); D.
Bodea, A. W\"urger, J. Low Temp. Phys. \textbf{136}, 39 (2004).

\bibitem {PFB}I. Ya. Polishchuk, P. Fulde, A. L. Burin, Y. Sereda, D.
Balamurugan, Journal of Low Temperature Physics, \textbf{140, }355 (2005).
\end{thebibliography}
\end{document}